\documentclass[manuscript]{aastex}
\usepackage{anysize}
\usepackage{hyperref} 
\usepackage{textcomp}
\usepackage{amsmath}
\shorttitle{PHOTOSPHERIC MAGNITUDE DIAGRAMS FOR SNe II}
\shortauthors{RODRIGUEZ ET AL.}

\newcommand\ubvriz{\mbox{$U\!BV\!RI\!Z$}}
\newcommand\bvri{\mbox{$BV\!RI$}}
\newcommand\vri{\mbox{$V\!RI$}}
\newcommand\bvi{\mbox{$BV\!I$}}
\newcommand\bi{\mbox{$B\!-\!I$}}
\newcommand\vi{\mbox{$V\!-\!I$}}

\begin{document}

\title{Photospheric Magnitude Diagrams for Type II Supernovae: A Promising Tool to Compute Distances}
\author{\'{O}smar Rodr\'{i}guez\altaffilmark{1}, Alejandro Clocchiatti\altaffilmark{2,1}, and Mario Hamuy\altaffilmark{3,1}}
\altaffiltext{1}{Millennium Institute of Astrophysics, Santiago, Chile}
\altaffiltext{2}{Instituto de Astrof\'isica, Facultad de F\'isica, Pontificia Universidad Cat\'olica de Chile, Casilla 306, Santiago 22, Chile}
\altaffiltext{3}{Departamento de Astronom\'ia, Universidad de Chile, Casilla 36-D, Santiago, Chile}

\begin{abstract}
We develop an empirical color-based standardization for Type II supernovae (SNe II), equivalent to the classical surface brightness method given in \citet{Wesselink1969}. We calibrate it with SNe II with host galaxy distance measured with Cepheids, and well-constrained shock breakout epoch and extinction due to the host galaxy. We estimate the reddening with an analysis of the $\bv$ versus $\vi$ color-color curves, similar to that of \citet{Natali_etal1994}. With four SNe II meeting the above requirements, we build a photospheric magnitude versus color diagram (similar to an HR diagram) with a dispersion of 0.29 mag. We also show that when using time since shock breakout instead of color as independent variable, the same standardization gives a dispersion of 0.09 mag. Moreover, we show that the above time-based standardization corresponds to the generalization of the standardized candle method of \citet{Hamuy_Pinto2002} for various epochs throughout the photospheric phase. To test the new tool, we construct Hubble diagrams to different subsamples of 50 low-redshift ($cz<10^4$ km s$^{-1}$) SNe II. For 13 SNe within the Hubble flow ($cz_{\text{\tiny CMB}}>3000$ km s$^{-1}$) and with well-constrained shock breakout epoch we obtain values of 68--69 km s$^{-1}$ Mpc$^{-1}$ for the Hubble constant, and an mean intrinsic scatter of 0.12 mag or 6\% in relative distances.
\end{abstract} 

\section{Introduction}
Core-collapse supernovae (SNe) are recognized as the final stage of stars with initial mass $\gtrsim$8 M$_\sun$ characterized by their great luminosities, comparable to the total luminosity of their host galaxies, and a sustained growth in size (an envelope ejection) due to the energy released by the gravitational collapse of the iron core \citep[e.g.,][]{Burrows2000}. Among these SNe, many show evidence for hydrogen (H) in their spectra, being historically classified as Type II \citep{Minkowski1941}. For this spectral type, \citet{Barbon1979} proposed a subclassification based on the shape of the light curve during the photospheric phase, i.e., the stage between maximum light and the transition to the radioactive tail: Type II ``plateau'' (II-P) for those SNe that display a nearly constant optical luminosity during $\sim$100 days, and  Type II ``linear'' (II-L) for those that show a linearly declining luminosity. Two other subclasses have been introduced based on spectral characteristics: IIn for SNe that show narrow H emission lines \citep{Schlegel_1990}, and IIb for those that show H in early spectra that soon disappear \citep{Woosley_etal1987}. Finally, few SNe do not fit into the aforementioned subclasses, but show characteristics similar to SN 1987A \citep[e.g.,][]{Kleiser_etal2011,Taddia_etal2012,Pastorello_etal2012}. More recently, \citet{Anderson_etal2014} performed a photometric analysis of a sample of 116 SNe II. They did not find a clear separation between II-P and II-L subclasses, so they propose a continuous subclass, named simply ``II'', where each SN is characterized by the decline rate during the plateau phase. Core-collapse SNe are the most frequent of all SNe \citep[76\% belonging to this class,][]{Li_etal2011}, being SNe II (II-P and II-L) the most frequent among core-collapse events \citep[$\sim$55\%,][]{Smith_etal2011}.

The physics of radiative transfer during the photospheric phase of SNe II-P is the better understood and probably the easiest to model for SN theorists \citep[e.g.,][]{Eastman_etal1996,Dessart_Hillier2005}. The above is further strengthened by the clear identification of red supergiants as progenitors of SNe II-P, fundamental to consolidate the hypothesis of a core-collapse process \citep{Smartt2009}. All these characteristics (i.e., high luminosity, high relative frequency, and physical understanding of the phenomenon) make SNe II-P well-suited for in-depth study, and interesting for distance measurement. 

At present, thermonuclear (Ia) SNe are the more mature astrophysical tool to measure distances: they are $\sim$1--2 mag brighter than SNe II-P in the optical, and have been shown to display the highest degree of both photometric and spectroscopic homogeneity \citep{Li_etal2001}. Type Ia SNe are not perfect standard candles, but empirical calibrations have allowed us to standardize their luminosities with a dispersion  in absolute magnitude of $\sim$0.15--0.22 mag and therefore determine distances to their host galaxies with an unprecedented precision of $\sim$7--10\% \citep{Phillips1993,Hamuy_etal1996,Phillips_etal1999}.

First attempts to measure distances to SNe II assume that these phenomena emit as blackbodies \citep[e.g.,][]{Kirshner_Kwan1974}. However, the scattering-dominated atmospheres of SNe II are not well-suited for the simple blackbody approximation \citep{Wagoner1981}, and this becomes an important factor to consider in distance measurements. Proposed solutions come from theoretical approaches to SN II atmospheres by \citet{Eastman_etal1996} and \citet{Dessart_Hillier2005}. However, significant systematic discrepancies exist between these two sets of models, leading to $\sim$50\% differences in distances \citep{Jones_etal2009}.

To get around this problem, empirical methods have been developed to estimate distances to SNe II \citep[e.g.,][]{Hamuy_Pinto2002}, but they do not make full use of the SNe II-P simplified model physics. Therefore, on the one hand we have theory-based methods with conflicting results, and, on the other, empirical methods that work well, but do not make full use of the possibilities of the underlying simple model. 

The goal of this research is to take advantage of the simplified physics of SNe II to find improved ways of estimating distances.

We organize our work as follows: In \S \ref{observational_material} we describe the photometric and spectroscopic data. In \S \ref{tools_to_measure_distances} we develop the standard framework, and a standardization based on color is presented in \S \ref{photospheric_magnitude}. In \S \ref{crucial_parameters} we discuss some crucial parameters involved in our analysis, with emphasis in the development of a method to measure host galaxy extinction in \S \ref{host_galaxy_extinction}. In \S \ref{analysis} we show our results: extinction measurements, the color-based standardization, and an alternative standardization based on time since shock breakout. In \S \ref{discussion} we discuss some aspects of the classical surface brightness method, and the next refinement steps for the method proposed in this work. We present our conclusions in \S \ref{conclusions}.

\section{Observational Material}\label{observational_material}
We base our work in data obtained in the course of four different systematic SN follow-up programs: the Cerro Tololo SN program (PIs: Phillips and Suntzeff, 1986--2003), the Cal\'an/Tololo SN survey (PI: Hamuy, 1989--1993), the Supernova Optical and Infrared Survey (SOIRS; PI: Hamuy, 1999--2000), and the Carnegie Type II Supernova Survey (CATS; PI: Hamuy, 2002--2003). The data were obtained with telescopes from Cerro Tololo Inter-American Observatory (CTIO), Las Campanas Observatory (LCO), the European Southern Observatory (ESO) in La Silla, and Steward Observatory (SO). Details on the telescope-instrument combinations are provided in \citet{Jones_2008}. The four programs obtained optical (and some IR) photometry and spectroscopy for nearly 100 SNe of all types, 47 of which belong to the Type II class (excluding IIn, IIb, and 1987A-like events). All of the optical data and the detailed description of the instruments used and the data reduction methods are being prepared for publication. Next we briefly summarize the general techniques used to obtain the photometric and spectroscopic data.

\subsection{Photometric data}
The photometric observations were carried out with the Johnson--Kron--Cousins--Hamuy $\ubvriz$ broadband filters \citep{Johnson_etal1966,Cousins1971,Hamuy_etal2001}.
The images were processed with IRAF\footnote{IRAF is distributed by the National Optical Astronomy Observatory, which is operated by the Association of Universities for Research in Astronomy (AURA) under cooperative agreement with the National Science Foundation.} through bias subtraction and flat-field correction. All of them were further processed through host galaxy subtraction using template images of the host galaxies. Photometric sequences were established around each SN based on observations of Landolt and Hamuy flux-standard stars \citep{Landolt_1992,Hamuy_etal1992,Hamuy_etal1994}. The photometry of all SNe was performed differentially with respect to the local sequence on the galaxy-subtracted images. The transformation of instrumental magnitudes into the standard system was done by taking into account a linear color term and a zero point.

\subsection{Spectroscopic data}
Low resolution ($R\sim1000$) optical spectra (wavelength range $\sim$ 3200--10000 \AA) were taken for each SN followed by comparison lamp spectra taken at the same position in the sky, and 2--3 flux-standard stars per night from the list of \citet{Hamuy_etal1992,Hamuy_etal1994}. Most of the spectra were taken with the slit along the parallactic angle \citep{Filippenko_1982}. The reductions were performed with IRAF and consisted of bias subtraction, flat-field correction, 1D-spectrum extraction and sky subtraction, wavelength calibration, and flux calibration.

\subsection{Sample of supernovae}
Among the 47 SNe II observed in the course of the four surveys, a subset of 33 objects comply with the requirements of 1) having photometry in $\bvi$ bands covering the photospheric phase; 2) having at least one measurement of expansion velocity; 3) having redshifts lower than $10^4$ km s$^{-1}$ (see subsection 3.3). To this sample, we added 17 SNe II from the literature\footnote{Spectroscopy for most of these SNe is available via WISeREP \citep[\url{http://www.weizmann.ac.il/astrophysics/wiserep/};][]{Yaron_GalYam2012}.}: SN 1999gi, SN 2001X, SN 2003Z, SN 2003gd, SN 2004A, SN 2004dj, SN 2004et, SN 2005ay, SN 2005cs, SN 2008in, SN 2009N, SN 2009bw, SN 2009js, SN 2009md, SN 2012A, SN 2012aw, and SN 2013ej. Spectroscopy of SN 2008in and SN 2009N is from the Carnegie Supernova Project \citep[CSP;][]{Hamuy_etal2006}. We also added the spectroscopy of SN 1999em published by \citet{Leonard_etal2002a}, and the spectroscopy of SN 2013ej obtained by C. Buil\footnote{\url{http://www.astrosurf.com/buil/}}. Table \ref{SN_sample} summarizes our final sample of 50 SNe II, which includes the name of the host galaxy, the heliocentric redshift, the reddening due to our Galaxy \citep{Schlafly_Finkbeiner2011}, the old SN subclassification (II-P or II-L) based on the $B$-band decline rate ($\beta_{100}^B$) criterion given in \citet{Patat_etal1994}, and the survey or references for the data.

\begin{table}[p]\scriptsize
    \caption{SN II Sample}
    \label{SN_sample}
    \begin{minipage}{\columnwidth}
      \centering
      \renewcommand{\arraystretch}{0.667}
      \begin{tabular}{l l c c c c c}\hline\hline
    SN Name & Host Galaxy & $cz_{\text{helio}}$\tablenotemark{a} & Source\tablenotemark{b} & $E(\bv)_{\text{Gal}}$\tablenotemark{c} & Type & References \\
             &             &   (km s$^{-1}$)                      &                         & (mag)                                  &      &            \\\hline
  1991al  & anon                       & 4572       & H01  & 0.0438 & II-L &  1                      \\
  1992af  & ESO 340--G38               & 5517       & here & 0.0449 & II-P &  1                      \\
  1992ba  & NGC 2082                   & 1052       & here & 0.0502 & II-P &  1                      \\
  1993A   & 2MASX J07391822--6203095   & 8707       & here & 0.1515 & II-P &  1                      \\
  1993S   & 2MASX J22522390--4018432   & 9623       & here & 0.0129 & II-L &  1                      \\
  1999br  & NGC 4900                   & \phn874    & here & 0.0203 & II-P &  2                      \\
  1999ca  & NGC 3120                   & 2791       & NED  & 0.0939 & II-L &  2                      \\
  1999cr  & ESO 576--G34               & 6050       & here & 0.0846 & II-P &  2                      \\
  1999eg  & IC 1861                    & 6703       & NED  & 0.1014 & II-L &  2                      \\
  1999em  & NGC 1637                   & \phn800    & L02a & 0.0349 & II-P &  2, 3                   \\
  1999gi  & NGC 3184                   & \phn503    & here & 0.0144 & II-P &  4                      \\
  2001X   & NGC 5921                   & 1480       & NED  & 0.0342 & II-P &  5, 6\tablenotemark{d}  \\
  2002ew  & NEAT J205430.50--000822.0  & 8834       & here & 0.0882 & II-L &  7                      \\
  2002gd  & NGC 7537                   & 2453       & here & 0.0575 & II-P &  7                      \\
  2002gw  & NGC  992                   & 3059       & here & 0.0165 & II-P &  7                      \\
  2002hj  & NPM1G +04.0097             & 6994       & here & 0.0992 & II-P &  7                      \\
  2002hx  & 2MASX J08273975--1446551   & 9214       & here & 0.0460 & II-P &  7                      \\
  2003B   & NGC 1097                   & 1058       & here & 0.0232 & II-P &  7                      \\
  2003E   & MCG --04--12--004          & 4400       & here & 0.0417 & II-P &  7                      \\
  2003T   & UGC 4864                   & 8368       & NED  & 0.0272 & II-P &  7                      \\
  2003Z   & NGC 2742                   & 1289       & NED  & 0.0335 & II-P &  8                      \\
  2003bl  & NGC 5374                   & 4211       & here & 0.0232 & II-P &  7                      \\
  2003bn  & 2MASX J10023529--2110531   & 3813       & here & 0.0563 & II-P &  7                      \\
  2003ci  & UGC 6212                   & 8967       & here & 0.0508 & II-L &  7                      \\
  2003cn  & IC 849                     & 5430       & NED  & 0.0184 & II-P &  7                      \\
  2003ef  & NGC 4708                   & 4093       & here & 0.0395 & II-P &  7                      \\
  2003ej  & UGC 7820                   & 5056       & here & 0.0167 & II-L &  7                      \\
  2003fb  & UGC 11522                  & 5081       & here & 0.1559 & II-P &  7                      \\
  2003gd  & NGC 628                    & \phn657    & NED  & 0.0604 & II-P &  7, 9                   \\
  2003hg  & NGC 7771                   & 4186       & here & 0.0631 & II-P &  7                      \\
  2003hk  & NGC 1085                   & 6880       & here & 0.0313 & II-L &  7                      \\
  2003hl  & NGC 772                    & 2123       & here & 0.0624 & II-P &  7                      \\
  2003hn  & NGC 1448                   & 1168       & NED  & 0.0122 & II-P &  7                      \\
  2003ho  & ESO 235--G58               & 4091       & here & 0.0339 & II-P &  7                      \\
  2003ib  & MCG --04--48--015          & 7442       & NED  & 0.0418 & II-P &  7                      \\
  2003ip  & UGC 327                    & 5398       & NED  & 0.0564 & II-L &  7                      \\
  2003iq  & NGC 772                    & 2331       & here & 0.0624 & II-P &  7                      \\
  2004A   & NGC 6207                   & \phn852    & NED  & 0.0131 & II-P & 10, 11, 12              \\
  2004dj  & NGC 2403                   & \phn221    & V06  & 0.0344 & II-P & 13, 14                  \\
  2004et  & NGC 6946                   & \phn\phn40 & NED  & 0.2937 & II-P & 15, 16                  \\
  2005ay  & NGC 3938                   & \phn809    & NED  & 0.0183 & II-P & 17, 6\tablenotemark{d}  \\ 
  2005cs  & NGC 5194                   & \phn463    & NED  & 0.0308 & II-P & 18, 19                  \\
  2008in  & NGC 4303                   & 1566       & NED  & 0.0192 & II-P & 20, 21                  \\
  2009N   & NGC 4487                   & 1036       & NED  & 0.0179 & II-P & 22, 21                  \\
  2009bw  & UGC 2890                   & 1155       & NED  & 0.1976 & II-P & 23\tablenotemark{d}     \\
  2009js  & NGC 918                    & 1507       & NED  & 0.3031 & II-P & 24                      \\
  2009md  & NGC 3389                   & 1308       & NED  & 0.0233 & II-P & 25, 26\tablenotemark{d} \\
  2012A   & NGC 3239                   & \phn753    & NED  & 0.0274 & II-P & 27\tablenotemark{d}     \\
  2012aw  & NGC 3351                   & \phn778    & NED  & 0.0239 & II-P & 28, 29                  \\ 
  2013ej  & NGC 628                    & \phn657    & NED  & 0.0597 & II-L & 30, 31                  \\\hline\\[-1.2cm]
      \tablenotetext{a}{\scriptsize Adopted heliocentric SN redshift.}
      \tablenotetext{b}{\scriptsize Source of SN redshift---NED: NASA/IPAC Extragalactic Database (\url{http://ned.ipac.caltech.edu/}); H01: \citet{Hamuy2001}; L02a: \citet{Leonard_etal2002a} V06: \citet{Vinko_etal2006}; here: this work.}
      \tablenotetext{c}{\scriptsize Galactic reddenings from \citet{Schlafly_Finkbeiner2011}.}
      \tablenotetext{d}{\scriptsize Expansion velocities measured in the reference.}
      \tablerefs{\scriptsize (1)  Cal\'an/Tololo SN survey;
                             (2)  SOIRS;
                             (3)  \citealt{Leonard_etal2002a};
                             (4)  \citealt{Leonard_etal2002b};
                             (5)  \citealt{Tsvetkov2006};
                             (6)  \citealt{Poznanski_etal2009};
                             (7)  CATS;
                             (8)  \citealt{Spiro_etal2014};
                             (9)  \citealt{VanDyk_etal2003};
                             (10) \citealt{Tsvetkov2008};
                             (11) \citealt{Maguire_etal2010};
                             (12) \citealt{Hendry_etal2006};
                             (13) \citealt{Vinko_etal2006};
                             (14) \citealt{Tsvetkov_etal2008};
                             (15) \citealt{Sahu_etal2006};
                             (16) \citealt{Misra_etal2007};
                             (17) \citealt{Tsvetkov_etal2006};
                             (18) \citealt{Pastorello_etal2006};
                             (19) \citealt{Pastorello_etal2009};
                             (20) \citealt{Roy_etal2011};
                             (21) CSP;
                             (22) \citealt{Takats_etal2014};
                             (23) \citealt{Inserra_etal2012};
                             (24) \citealt{Gandhi_etal2013};
                             (25) \citealt{Fraser_etal2011};
                             (26) \citealt{Bose_Kumar2014};
                             (27) \citealt{Tomasella_etal2013};
                             (28) \citealt{Bose_etal2013};
                             (29) \citealt{DallOra_etal2014};
                             (30) \citealt{Richmond2014};
                             (31) C. Buil.}
      \end{tabular}
    \end{minipage}
\end{table}

\section{Tools to Measure Distances}\label{tools_to_measure_distances}
For the sake of simplicity, we assume a SN with a spherically symmetric expanding photosphere that radiates isotropically. We choose to work with photon flux instead of energy flux due to the fact that observations are obtained with photon detectors.

\subsection{Standard framework}\label{standard_framework}
In the relation between distance and photon flux, for the general case of a SN in a galaxy at redshift $z$ within the Hubble flow, we need to take into account the effects of the Universe expansion. Photon wavelengths are stretched and the time is dilated, both by a factor $(1+z)$. Using the $i$  subindex to indicate quantities at a given time $t_i$, and distinguishing between the observer's frame (unprimed) and the SN rest-frame (primed), we will have
\begin{equation}\label{lambda_z}
 \lambda'=\lambda/(1+z),
\end{equation}
and
\begin{equation}\label{high_z}
 n_{\lambda,i} = \mathcal{N}_{\lambda',i} \left(\frac{R'_{\text{ph},i}}{D_L}\right)^2 10^{-0.4(A_h(\lambda')+A_G(\lambda))}.
\end{equation}
Here, $n_\lambda$ is the monochromatic photon flux (in photons cm$^{-2}$ s$^{-1}$ \AA$^{-1}$) measured by a local observer from a SN at luminosity distance $D_L$, with an emergent photon flux $\mathcal{N}_{\lambda'}$ emitted from the photosphere with radius $R'_{\text{ph}}$, affected by dust extinction in the host galaxy $A_h(\lambda')$ and our Galaxy $A_G(\lambda)$.

In terms of broadband fluxes, Equation \ref{high_z} can be expressed as
\begin{align}\label{PRE_PMM1}
 m_{\overline\lambda,i} & \equiv -2.5\log\int{d\lambda S_{\overline\lambda}(\lambda) n_{\lambda,i}} + \text{ZP}_{\overline\lambda}\nonumber\\
                        & = 5\log{\left(\frac{D_L}{R'_{\text{ph},i}}\right)}+\widetilde{\mathcal{M}}_{\overline\lambda,i}(A_G,z,A_h),
\end{align}
where $m_{\overline\lambda}$ is the apparent magnitude in a photometric band with central wavelength $\overline\lambda$ and band transmission function $S_{\overline\lambda}(\lambda)$, and 
\begin{equation}\label{dirty_Mph}
 \widetilde{\mathcal{M}}_{\overline\lambda,i}(A_G,z,A_h) \equiv -2.5\log\int{d\lambda S_{\overline\lambda}(\lambda) \mathcal{N}_{\lambda',i}10^{-0.4(A_h(\lambda')+A_G(\lambda))}} + \text{ZP}_{\overline\lambda}.
\end{equation}
ZP$_{\overline\lambda}$ is the zero point of the magnitude scale, which is determined forcing the synthetic magnitude of a star with spectrophotometry $n_\lambda^{\text{star}}$ to match its apparent magnitude $m_{\overline\lambda}^{\text{star}}$, i.e.,
\begin{equation}\label{ZP}
 m_{\overline\lambda}^{\text{star}} =  -2.5\log\int{d\lambda S_{\overline\lambda}(\lambda)n_{\lambda}^{\text{star}}} + \text{ZP}_{\overline\lambda}.
\end{equation}
We can write Equation \ref{PRE_PMM1} as
\begin{equation}\label{PRE_PMM2}
 \mathcal{M}_{\overline\lambda,i}= m^{\text{corr}}_{\overline\lambda,i}-A_h\left(\overline\lambda\right)-5\log{\left(\frac{D_L}{R'_{\text{ph},i}}\right)},
\end{equation}
\begin{equation}
 m^{\text{corr}}_{\overline\lambda,i}\equiv m_{\overline\lambda,i}-A_G(\overline\lambda)-K_i(\overline\lambda),
\end{equation}
where $\mathcal{M}_{\overline\lambda,i}\equiv\widetilde{\mathcal{M}}_{\overline\lambda,i}(A_{G}=z=A_{h}=0)$, $A_{G}(\overline\lambda)$ and $A_{h}(\overline\lambda)$ are the Galactic and host galaxy broadband extinction respectively, and 
\begin{equation}\label{Ki}
 K_i(\overline\lambda)\equiv-2.5\log(1+z)+2.5\log\left(\frac{\int{d\lambda S_{\overline\lambda}(\lambda)\mathcal{N}_{\lambda,i}10^{-0.4A_h(\lambda)}}}{\int{d\lambda' S_{\overline\lambda}(\lambda)\mathcal{N}_{\lambda',i}10^{-0.4A_h(\lambda')}}}\right)
\end{equation}
is the $K$-term \citep[cf.][]{Schneider_etal1983}.

\subsection{Photospheric magnitude}\label{photospheric_magnitude}
We note that Equation \ref{PRE_PMM2} is similar to the definition of distance modulus. $M_{\overline\lambda}$ is defined as the magnitude seen at $10$ pc from the source. Similarly, Equation \ref{PRE_PMM2} relates the apparent magnitude with $\mathcal{M}_{\overline\lambda}$, a quantity not at $10$ pc but just above the photosphere. We will call it photospheric magnitude. Absolute and photospheric magnitude are related by
\begin{equation}\label{Mph_M}
 \mathcal{M}_{\overline\lambda,i}=M_{\overline\lambda,i}-\mathcal{R}_i,
\end{equation}
\begin{equation}
 M_{\overline\lambda,i}= m^{\text{corr}}_{\overline\lambda,i}-A_h\left(\overline\lambda\right)-\mu,
\end{equation}
where $\mu$ is the distance modulus, and
\begin{equation}
 \mathcal{R}_i\equiv5\log{\left(\frac{10\text{ pc}}{R'_{\text{ph},i}}\right)}.
\end{equation}

\subsection{Surface brightness method}\label{surface_brightness_method}
The photospheric magnitude in Equation \ref{PRE_PMM2} can also be interpreted as an indicator of surface brightness, which in a magnitude scale is defined as
\begin{equation}\label{Sl}
 s_{\overline\lambda}=m^{\text{corr}}_{\overline\lambda}+5\log\theta
\end{equation}
\citep[e.g.,][]{Wesselink1969}. The photosphere's angular size $\theta$, used to calculate the area of the object in the sky, is also an indicator of $R'_{\text{ph}}/D$.

For Galactic cool giants, supergiants, and Cepheid variables a relation is found between the surface brightness and a suitably chosen color index, that can be used as a proxy for temperature \citep{Welch_1994,Fouque_Gieren1997,Kervella_etal2004b}. In the case of SNe II, atmosphere models show that the emergent flux, which depends on many parameters (e.g., chemical composition or density structure of the progenitor star), has an important dependence on temperature and, in short-wavelength bandpasses, density at the photosphere \citep{Eastman_etal1996}. So we expect a relation between photospheric magnitude and a color index.

This kind of relation allows us to infer angular sizes using only photometric information and, comparing with the physical size of the object estimated from the expansion, calculate distances. This method is known as the surface brightness method \citep[SBM;][]{Barnes_Evans1976}. In order to obtain distances with the SBM we need to measure the physical radius and calibrate the specific color-magnitude relation for SNe II.

\subsubsection{Physical radius}\label{physical_radius}
The gravitational binding energy of a SN II progenitor ($\sim10^{49}$ erg) is much smaller than the expansion kinetic energy of the ejecta ($\sim10^{51}$ erg), and the mass of the interstellar and circumstellar matter swept up by the SN in the first few months is also much smaller than the ejected mass, so any possible deceleration is negligible and therefore the envelope undergoes free expansion.

The physical radius of the photosphere under such conditions is given by
\begin{equation}
 R_{\text{ph}} = R(m_{\text{ph}},t)\approx v(m_{\text{ph}})(t-t_0)+R_0(m_{\text{ph}}),
\end{equation}
where $m_{\text{ph}}$ is the Lagrangian mass coordinate of the material instantaneously at the photosphere moving with velocity $v(m_{\text{ph}})$ (hereafter photospheric velocity, $v_{\text{ph}}$), $t_0$ is the shock breakout epoch, and $R_0(m_{\text{ph}})$ is the radius of the SN at $t_0$.
Due to the high expansion velocities ($\sim10^4$ km s$^{-1}$), even for a large progenitor with $R_0\sim5\times10^{13}$ cm ($\sim700$ R$_\sun$), the initial radius is only $\sim10\%$ of the photospheric radius at five days since shock breakout, so $R_0$ can be neglected after a few days. Then, we can estimate the radius of the photosphere for all but the first few days with the simple expression (distinguishing between the observer's frame and the SN rest-frame)
\begin{equation}\label{Rph}
 R'_{\text{ph}} =v'_{\text{ph}}\frac{t-t_0}{1+z}.
\end{equation}

The photospheric velocity can be estimated from absorption minima of P Cygni profiles seen in a spectrum \citep{Kirshner_Kwan1974}. Traditionally, weak lines (e.g., \ion{Fe}{2} lines) are used for this purpose \citep[e.g.,][]{Schmidt_etal1992}. This is based under the assumption that those lines are optically thin above the photosphere, being better indicators of the photospheric velocity than strong, optically thick lines (e.g., Balmer lines) formed well above the photosphere \citep{Leonard_etal2002a}.

\subsubsection{Calibration of the photospheric magnitude}
Methods to measure distances to SNe II based on atmosphere models, as the expanding photosphere method \citep[EPM;][]{Kirshner_Kwan1974,Schmidt_etal1992}, have a typical uncertainty of 15$\%$ in distance \citep{Jones_etal2009}. Both the EPM and the SBM are based on the Baade's (1926) method, so the uncertainty of a SBM distance  will not be very different to the typical EPM uncertainty if we calibrate the color--photospheric magnitude relation with SN II atmosphere models.

In this work, however, we will obtain the photospheric magnitude calibration empirically. A calibration based on Equation \ref{Sl} is not possible because we cannot measure the SN angular size directly with the current optical instrumentation. Therefore, for the photospheric magnitude calibration, paying attention to Equation \ref{PRE_PMM2}, we need SNe in galaxies with well-determined distances. Also, for calibration and distance measurements, we need a precise knowledge of the SN shock breakout epoch and host galaxy extinction.

To measure expansion velocities and perform the $K$ correction we need to know the SN redshift. For 25 SNe we estimate this quantity measuring the peak wavelength of the narrow emission lines resulting from the superposed \ion{H}{2} region \citep[e.g.,][]{Leonard_etal2002b}. For the other 22 SNe the aforementioned emission line is not detected, so we adopt the redshift of the host galaxy nucleus, while for SN 1991al, SN 1999em, and SN 2004dj we adopt the values from \citet{Hamuy2001}, \citet{Leonard_etal2002a}, and \citet{Vinko_etal2006} respectively.
 
\citet{Schmidt_etal1994} discussed that $K$ corrections become significant for SNe at redshifts greater than $10^4$ km s$^{-1}$. Then, for the sake of simplicity in the first step to test the reliability of the photospheric magnitude, we use only SNe with $cz<10^4$ km s$^{-1}$ to avoid performing the $K$ correction. However, we will keep the first term of the $K$-term (Equation \ref{Ki}) to take into account the dispersion of up to $\sim0.04$ mag that introduces this term.

\section{Crucial Parameters}\label{crucial_parameters}

\subsection{Host galaxy extinction: the C3 method}\label{host_galaxy_extinction}
The host galaxy extinction can be estimated both from spectroscopic and photometric data. The main spectroscopic method is based on the comparison between an observed spectrum and SN II atmosphere model spectra, where the fitting parameters are the amount of reddening and the photospheric temperature \citep[e.g.][]{Dessart_Hillier2006,Dessart_etal2008}. The main photometric method for SNe II-P assumes that, due to the H recombination nature of their photospheres, they should evolve from a hot initial stage to one of constant photospheric temperature, reaching the same color at a certain epoch \citep[e.g.][]{Olivares_etal2010}. In both cases, differences between an observed SN and models, or other SNe consistent with zero reddening, are attributed to a color excess $E(C)$. The latter is associated to the host galaxy extinction $A(\overline\lambda)$ by means of the extinction-to-reddening ratio,
\begin{equation}\label{A_EC_rel}
 R_{\overline\lambda,C} = A(\overline\lambda)/E(C),
\end{equation}
where $R_{\overline\lambda,C}$ can be derived from a Cardelli reddening law \citep[hereafter CCM;][]{Cardelli_etal1989, ODonnell1994} as function of $R_V\equiv A_V/E(\bv)$.

The main problem to estimate reddening with the aforementioned photometric method is to select the time when SNe reach the same color (for which we also need to know the shock breakout epoch). To avoid these problems, we will use color instead of time to express the SN color evolution, i.e., we will use color-color curves (C3) to estimate reddening. 

With data from the $\{\bvri\}$ filter set, it is possible to define three independent color indexes. In this work we choose to use $\bv$, $\vr$ and $\vi$. They are related to their intrinsic values (hereafter marked with zero subindex) through
\begin{eqnarray}
    \bv&=&\left(\bv\right)_0+E\left(\bv\right),\nonumber\\
    \vr&=&\left(\vr\right)_0+E\left(\vr\right),\\
    \vi&=&\left(\vi\right)_0+E\left(\vi\right),\nonumber
\end{eqnarray}
\begin{figure}[p]
  \centering
  \includegraphics[width=\columnwidth]{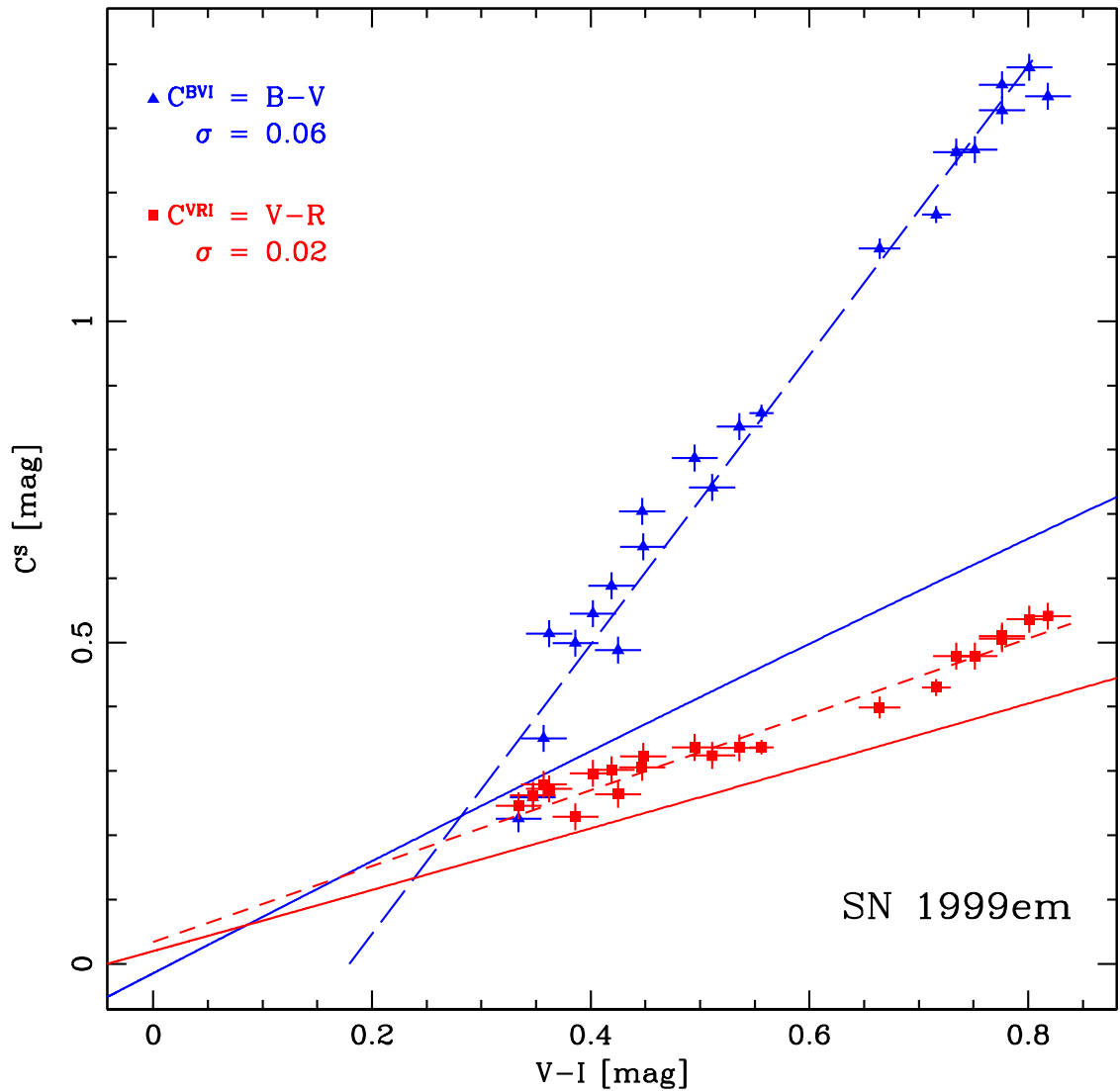}
  \caption{Color-color diagram showing the C3 of a blackbody (solid lines) and of SN 1999em (dashed lines) during the photospheric phase. Differences between them are due to the SN not being a perfect blackbody, to the presence of spectral lines and also to the reddening that affects the SN. Line blanketing makes the difference stronger in the $B$-band. The linear fit is more accurate for the $\vri$ points ($\sigma=0.02$) than for the $\bvi$ points ($\sigma=0.06$).}
  \label{1999em}
\end{figure}
where $E(\overline\lambda_1-\overline\lambda_2)=A(\overline\lambda_1)-A(\overline\lambda_2)$ are the color excesses. For a blackbody there is a very approximate linear relation between any two color indexes because both of them are approximated functions of $1/T$. We expect that SNe II will hold a similar relation during their photospheric phase. There will be differences, though, because SNe II are not perfect blackbodies, have a spectral diversity, and are also affected by reddening. Figure \ref{1999em} shows the $\bv$ versus $\vi$ and $\vr$ versus $\vi$ C3 for a blackbody (solid lines) and the data of SN 1999em in the photospheric phase (points). We see that the $\vr$ versus $\vi$ points have a linear behavior (short-dashed line) meaning that the blackbody assumption is fairly good. The presence of spectral peculiarities causes differences between observed and blackbody C3. This is more evident in color index $\bv$ (long-dashed line), where line blanketing generates a significant difference with a blackbody. Then, for the case of an unreddened SN, we will write the $\vr$ versus $\vi$ C3 as, approximately,
\begin{equation}
 \left(\vr\right)_0=n^{_{V\!RI}}_0+m^{_{V\!RI}}_0\left(\vi\right)_0\label{VR_0}.
\end{equation}
For the $\bv$ versus $\vi$ C3 we will write the same form, although it will be a less accurate approximation,
\begin{equation}
 \left(\bv\right)_0=n^{_{BV\!I}}_0+m^{_{BV\!I}}_0\left(\vi\right)_0\label{BV_0}.
\end{equation}
If there is extinction, points from Equations \ref{VR_0} and \ref{BV_0} will be moved in a color-color diagram (as Figure \ref{1999em}) to the right by the color excess $E(\vi)$, and upward by the color excesses $E(\vr)$ and $E(\bv)$ respectively, i.e., they will be shifted in the color-color diagram respectively by the reddening vectors
\begin{eqnarray}
 (E(\vi),E(\vr))\equiv\textbf{\textit{E}}_{V\!RI}\label{vecE},\\
 (E(\vi),E(\bv))\equiv\textbf{\textit{E}}_{BV\!I}\label{vecF}.
\end{eqnarray}
\begin{figure}[p]
  \centering
  \includegraphics[width=\columnwidth]{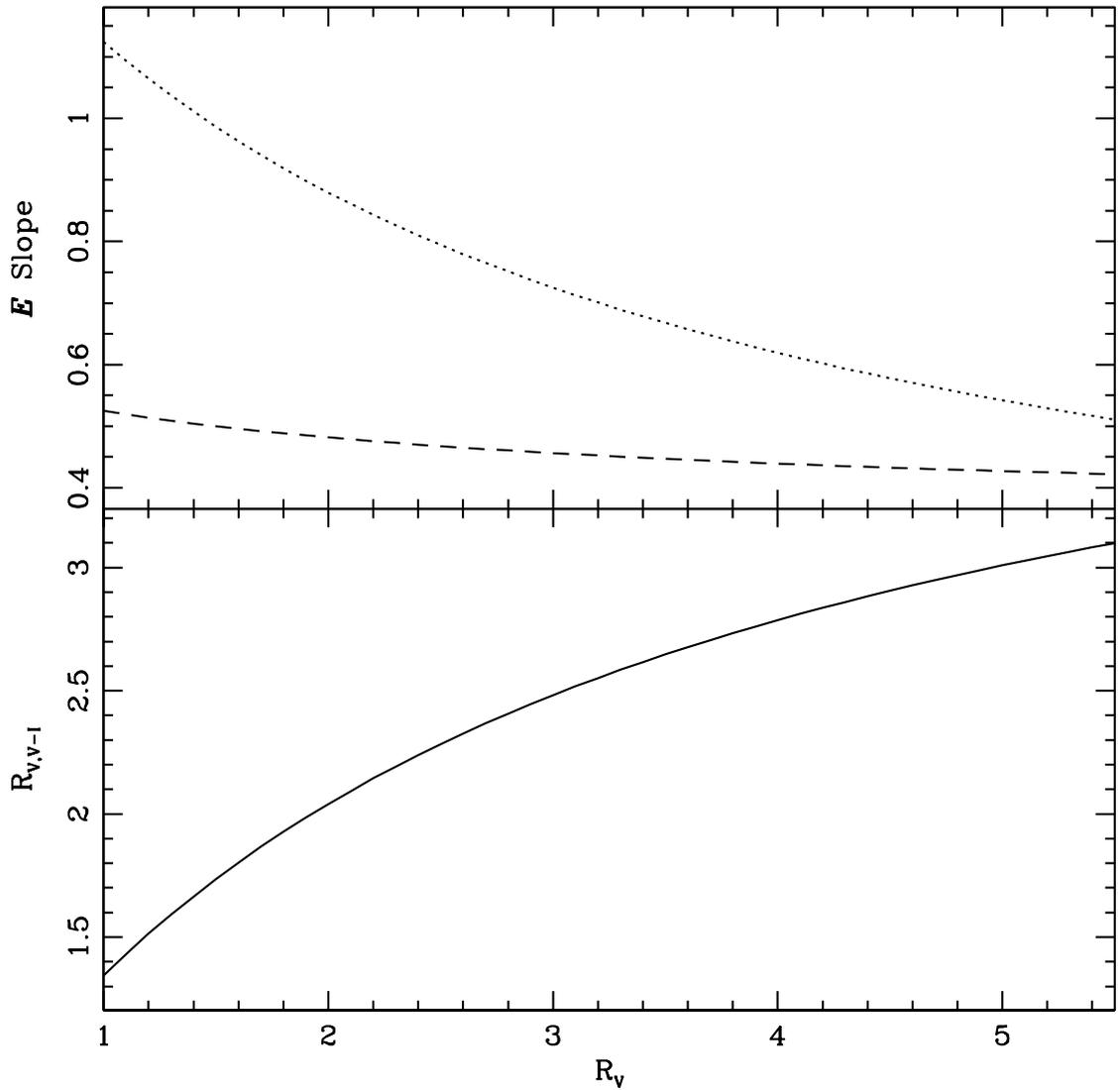}
  \caption{Dependence with $R_V$ of $m^{_{V\!RI}}_{\textbf{\textit{E}}}$, $m^{_{BV\!I}}_{\textbf{\textit{E}}}$ (top, dashed and dotted line respectively), and $R_{V,V\!-I}=A_V/E(\vi)$ (bottom). The curves are computed with SNe II spectra during the photospheric phase and the CCM reddening law.}
  \label{REDD_LAW_vs_Rv}
\end{figure}
Since the color excesses are proportional to $A_V$, the length of $\textbf{\textit{E}}_{V\!RI}$ and $\textbf{\textit{E}}_{BV\!I}$ are proportional to the amount of extinction, and their slopes 
\begin{equation}
 m^{_{V\!RI}}_{\textbf{\textit{E}}}\equiv \frac{E(\vr)}{E(\vi)},
\end{equation}
and
\begin{equation}
 m^{_{BV\!I}}_{\textbf{\textit{E}}}\equiv \frac{E(\bv)}{E(\vi)},
\end{equation}
are constants for a given $R_V$. Top of Figure \ref{REDD_LAW_vs_Rv} shows the dependence of $m^{_{V\!RI}}_{\textbf{\textit{E}}}$ and $m^{_{BV\!I}}_{\textbf{\textit{E}}}$ on $R_V$, both computed with SNe II spectra during the photospheric phase and the $R_V$-parametrization of $A_{\overline\lambda}/A_V$ given by the CCM reddening law. Thus, the reddening vectors only modify the zero point of the C3 in the color-color diagram but not its slopes. So, the C3 for the SN affected by reddening can be expressed as
\begin{eqnarray}
 \vr= n^{_{V\!RI}}+m^{_{V\!RI}}_0(\vi)\label{VR},\\
 \bv= n^{_{BV\!I}}+m^{_{BV\!I}}_0(\vi)\label{BV}.
\end{eqnarray}
Figure \ref{3g} shows the effect of the reddening vector over an unreddened C3. The C3 is moved above or below the unreddened C3 depending on whether the slope of the reddening vector is larger or smaller than the slope of the C3.
\begin{figure}[p]
  \centering
  \includegraphics[width=\columnwidth]{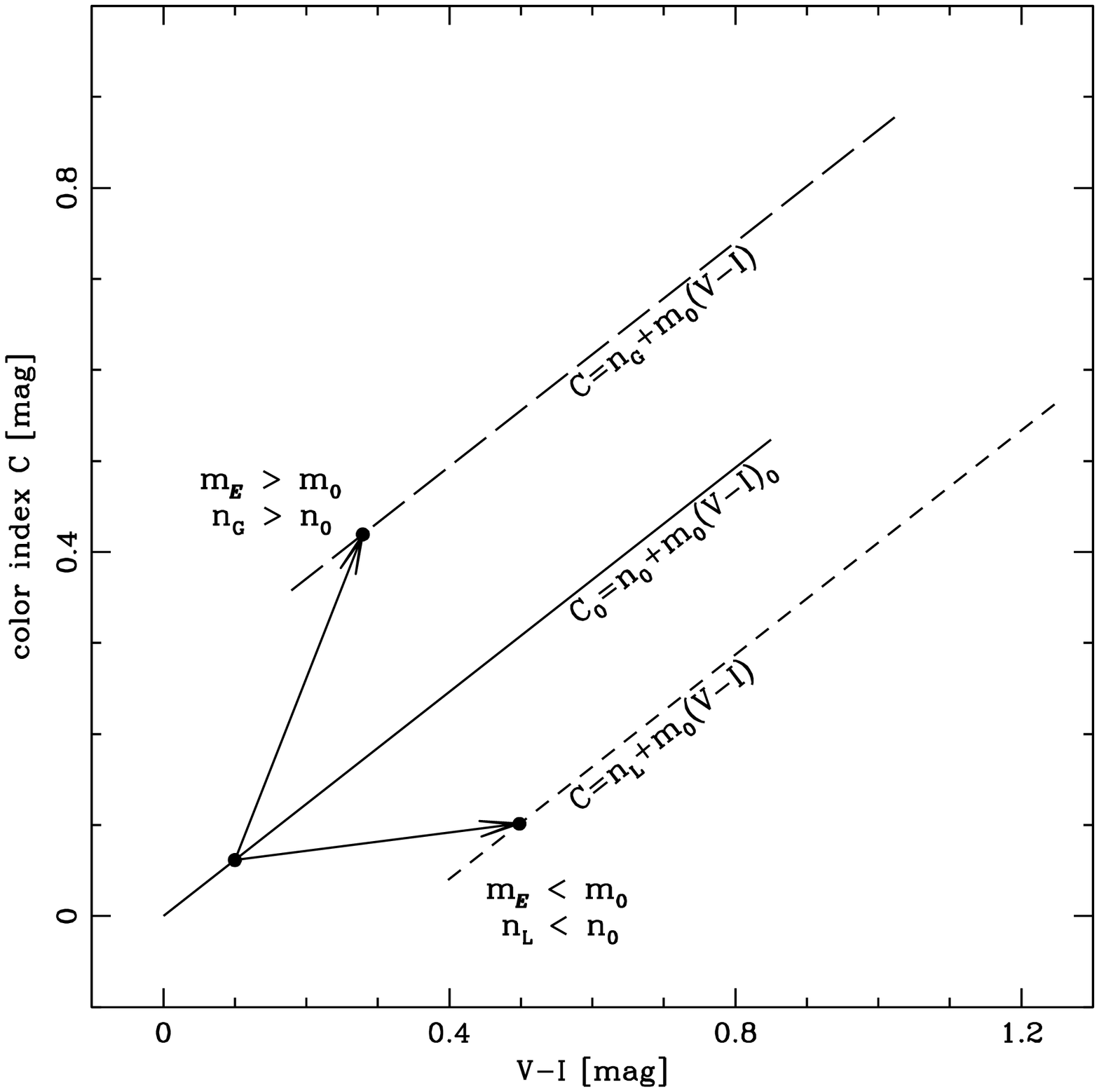}
  \caption{Example of the reddening effect on a linear C3. The unreddened C3 (solid line) is moved up if the slope of the reddening vector ($m_{\textbf{\textit{E}}}$) is larger than the C3 slope ($m_0$), producing a reddened C3 (long-dashed line) with a greater y-intercept ($n_G$). If $m_{\textbf{\textit{E}}}<m_0$, then the unreddened C3 is moved down, producing a reddened C3 (short-dashed line) with a lower y-intercept ($n_L$).}
  \label{3g}
\end{figure}

In general, with $\bvri$ photometric data, we can determine color excesses with the $\{\vri\}$ filter subset (Equation \ref{VR} minus Equation \ref{VR_0}) or with the $\{\bvi\}$ filter subset (Equation \ref{BV} minus Equation \ref{BV_0}), and the corresponding uncertainty via
\begin{equation}\label{E_S(V-I)}
 E_S(\vi)=\frac{n^{_S}-n^{_S}_0}{m^{_S}_{\textbf{\textit{E}}}(R_V)-m^{_S}_0},
\end{equation}
\begin{equation}\label{s_E}
 \sigma_{E_S(V\!-\!I)}=\frac{\sqrt{\sigma_{n^{_S}}^2+\sigma_{n^{_S}_0}^2+\sigma_{m^{_S}_0}^2E_S(\vi)^2}}{|m^{_S}_{\textbf{\textit{E}}}(R_V)-m^{_S}_0|},
\end{equation}
where $S=\{\vri\}$, $\{\bvi\}$. We see that the applicability of the method through Equation \ref{E_S(V-I)} depends strongly on the value of $|m^{_S}_{\textbf{\textit{E}}}(R_V)-m^{_S}_0|$: if the value is small, then the reddening vector moves points along the C3 and no reliable reddening determination is possible. The visual extinction is obtained through the extinction-to-reddening ratio (see bottom of Figure \ref{REDD_LAW_vs_Rv})
\begin{equation}\label{bV}
 R_{V,V\!-I}(R_V) = \frac{A_V}{E(\vi)}.
\end{equation}

We note that the method presented above is essentially the same method proposed by \citet{Natali_etal1994} to measure reddening for open clusters, which shows a linear relation between $\bi$ and $\bv$ color indexes \citep[for further discussion of the method, see][]{Munari_Carraro1996}. Hereafter we will refer to this method simply as the C3 method.

In order to compute color excesses with Equation \ref{E_S(V-I)}, we need to determine for each SN the unreddened C3 parameters $n^{_S}_0$ and $m^{_S}_0$ and the value of $R_V$. This task can be carried out in a simple way with the following two assumptions:

1. \textit{SNe II have a similar spectral evolution during the photospheric phase}.---
This is partly supported by results of SN II atmosphere models, which show that the emergent flux depends mainly on temperature \citep[][]{Eastman_etal1996,Jones_etal2009}. In that case, SNe II should have approximately the same intrinsic C3 with slope $m^{_S}_0$. As the C3 slope is not affected by reddening, the value of $m^{_S}_0$ does not depend on dust extinction or dust properties (i.e., reddening law and $R_V$). There could be, however, intrinsic differences due to the variations on the SN parameters, mainly those involved with the line profile formation (e.g., metallicity, density and expansion velocity profile).

2. \textit{Dust in SNe II sightlines is described by the same $R_V$}.---
SNe II tend to appear near \ion{H}{2} regions, so the value of $R_V$ to use should be representative for these regions. In that case, the slope of reddening vectors will be the same for all SNe. This allow us to identify the SN with the smallest reddening in a given SN set (see Figure \ref{3g}): that with the lowest value of $n^{_S}$ if the slope of the reddening vector is larger than $m^{_S}_0$, or that with the highest value of $n^{_S}$ if the slope of the reddening vector is lower than $m^{_S}_0$. If the SN found with the smallest reddening by this method is consistent with small reddening found by other methods, then we can set its y-intercept as $n^{_S}_0$.

\subsection{Shock breakout epoch}\label{shock_breakout_epoch}
The shock breakout epoch $t_0$ is one of the most important parameters in determining the actual stage of the expanding photosphere. This value is estimated as lying between the last nondetection $t_n$ and the first detection $t_f$ of the SN, where the first detection can be the discovery epoch $t_d$ or the time of a prediscovery detection if it is available. Defining $t_m$ as the midpoint between $t_n$ and $t_f$, and the uncertainty $\Delta t$ as half the time between $t_n$ and $t_f$, i.e.,
\begin{equation}
 t_m = (t_n+t_f)/2-t_d,
\end{equation}\begin{equation}
 \Delta t = (t_f-t_n)/2
\end{equation}
(where we choose to express time with respect to $t_d$), we can locate the shock breakout epoch in the range $t_m-\Delta t<t_0<t_m+\Delta t$ \citep[e.g.,][]{DAndrea_etal2010}. We use $\Delta$ instead of $\sigma$ to emphasize that the uncertainty is not gaussian but uniform, with a probability distribution described by a top hat function. 

If a SN has a $t_n$ that yields a very high $\Delta t$, or simply there is no nondetection report for the SN, then we use the shock breakout epoch obtained through the Supernova Identification code \citep[SNID;][]{Blondin_Tonry2007} given in \citet{Anderson_etal2014}. For some SNe we compare their earliest spectra with a library of SN spectra using the GELATO\footnote{\url{https://gelato.tng.iac.es}} tool \citep{Harutyunyan_etal2008}, then we constrain $t_0$ between the lowest and the highest value of $t_0$ estimated with the spectra comparison.

Table \ref{ladj} lists the midpoint epochs and the uncertainties (Column 7) for our SN set, with the respectrive references (Column 8).

\section{Analysis}\label{analysis}

\subsection{Application of the C3 method}\label{comparing_reddening_methods}
\begin{figure}[p]
  \centering
  \includegraphics[width=\columnwidth]{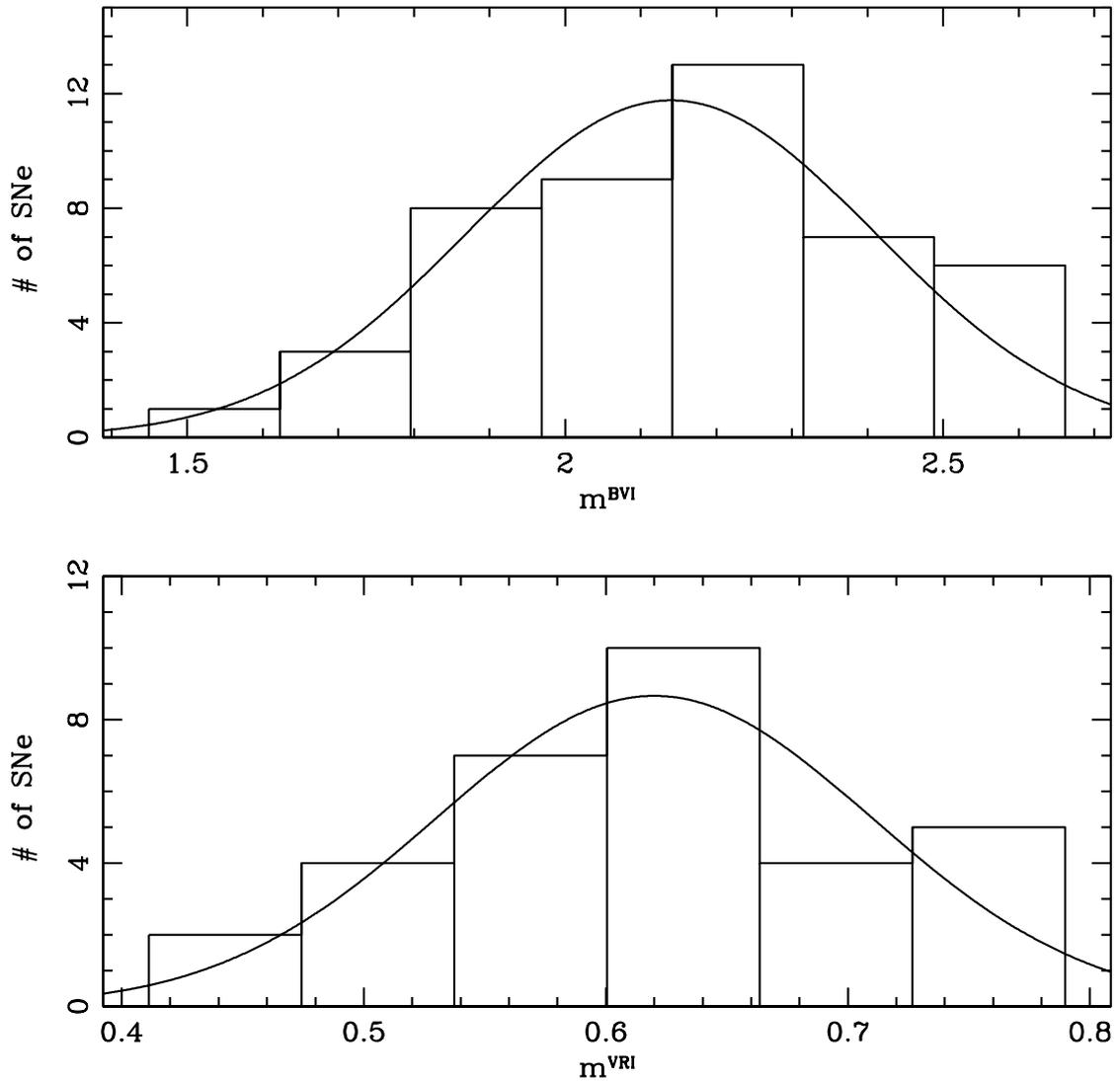}
  \caption{C3 slope distribution for SNe for the $\{\bvi\}$ (top) and $\{\vri\}$ (bottom) filter subsets, where the solid line is the gaussian distribution computed with the respective values for the mean and standard deviation.}
  \label{m_histo}
\end{figure}

Table \ref{ladj} lists the C3 slope values obtained for the SNe in the $\{\vri\}$ (Column 2) and $\{\bvi\}$ (Column 3) filter subsets. SN 2003gd and SN 2003ho do not describe clear straight lines in the color-color diagram because for these SNe we have only late-time photometry. So we discard them for the C3 intrinsic slope determination. We also rule out SN 1991al and SN 2002ew due to the poor sampling in the $B$ and $I$ band respectively. We obtain mean slopes of $\langle m^{_{BV\!I}}\rangle=2.14$ and $\langle m^{_{V\!RI}}\rangle=0.62$ with standard deviations of 0.27 and 0.09 respectively. In each filter subset the measurement errors for the C3 slopes are lower than the standard deviation, so this value represents the intrinsic diversity in the C3 slope value. Therefore we adopt $m^{_{BV\!I}}_0=2.14\pm0.27$ and $m^{_{V\!RI}}_0=0.62\pm0.09$. Figure \ref{m_histo} shows the distribution of C3 slopes.

The next step is to compare the value of the intrinsic C3 slope with the respective reddening vector slopes. For $R_V\geq1$, $|m^{_{V\!RI}}_{\textbf{\textit{E}}}-m^{_{V\!RI}}_0|\leq0.2$ (see top of Figure \ref{REDD_LAW_vs_Rv}). This means that in the case of a C3$_{V\!RI}$ the displacement produced by reddening is virtually along the same C3, so the C3 method for the $\{\vri\}$ filter subset is not suitable to measure reddening. For the $\{\bvi\}$ filter subset $|m^{_{BV\!I}}_{\textbf{\textit{E}}}-m^{_{BV\!I}}_0|>1$, so the C3 method for this filter subset (hereafter C3($\bvi$)) is very promising for measuring reddening.

\begin{table}[p]\scriptsize
    \caption{SN II Parameters}
    \label{ladj}
    \begin{minipage}{\columnwidth}
      \centering
      \renewcommand{\arraystretch}{0.667}
      \begin{tabular}{l c c c c c c l}\hline\hline
      SN Name &$m^{_{V\!RI}}(\sigma)$&$m^{_{BV\!I}}(\sigma)$&$n^{_{BV\!I}}(\sigma)$\tablenotemark{a}& $A_V$(spec)($\sigma$)\tablenotemark{b}  & $A_V(\bvi)(\sigma)$  & $t_m(\Delta t)$\tablenotemark{c}   & References\tablenotemark{d}  \\
              &            &            &      (mag)           &   (mag)                     &       (mag)           &   (days)                                     &                      \\\hline
      1991al  & 0.67(0.02) & 3.64(0.06) & \textminus0.57(0.07) & 0.33(0.16)                  &       \phs0.61(0.17)  &     \textminus15.2(\phn9.0)\tablenotemark{e} & I 5310, A14       \\
      1992af  & \nodata    & 2.05(0.04) & \textminus0.14(0.07) & 1.26(0.31)                  & \textminus0.15(0.13)  &     \textminus12.3(\phn6.0)\tablenotemark{e} & I 5554, A14       \\
      1992ba  & \nodata    & 2.18(0.02) & \textminus0.32(0.05) & 0.46(0.16)                  &       \phs0.16(0.10)  & \phn\textminus 6.5(\phn6.5)                  & I 5625, 5632      \\
      1993A   & \nodata    & 2.27(0.03) & \textminus0.29(0.04) & 0.08(0.31)                  &       \phs0.10(0.09)  & \phn\textminus 9.6(\phn9.6)                  & I 5693            \\
      1993S   & \nodata    & 2.47(0.04) & \textminus0.23(0.05) & \nodata                     &       \phs0.00(0.10)  & \phn\textminus 4.4(\phn4.0)\tablenotemark{e} & I 5812, A14       \\
      1999br  & 0.49(0.01) & 2.01(0.02) & \textminus0.58(0.04) & 0.26(0.16)                  &       \phs0.62(0.14)  & \phn\textminus 4.0(\phn4.0)                  & I 7141, 7143      \\
      1999ca  & 0.73(0.01) & 2.52(0.02) & \textminus0.31(0.05) & 0.17(0.31)                  &       \phs0.15(0.10)  &     \textminus19.6(\phn7.0)\tablenotemark{e} & I 7158, A14       \\
      1999cr  & 0.76(0.01) & 2.35(0.02) & \textminus0.55(0.04) & 0.51(0.31)                  &       \phs0.57(0.13)  & \phn\textminus 3.2(\phn7.0)\tablenotemark{e} & I 7210, A14       \\
      1999eg  & 0.71(0.02) & 2.56(0.04) & \textminus0.22(0.04) & \nodata                     & \textminus0.01(0.07)  & \phn\textminus 2.1(\phn2.1)\tablenotemark{e} & I 7275, H08       \\
      1999em  & 0.60(0.01) & 2.24(0.02) & \textminus0.28(0.03) & 0.33(0.16)                  &       \phs0.09(0.06)  & \phn\textminus 4.5(\phn4.5)                  & I 7294            \\
      1999gi  & 0.57(0.01) & 2.24(0.03) & \textminus0.79(0.05) & 0.57(0.16)                  &       \phs0.99(0.21)  & \phn\textminus 3.3(\phn3.3)                  & I 7329, 7334      \\
      2001X   & 0.62(0.03) & 2.34(0.11) & \textminus0.38(0.06) & \nodata                     &       \phs0.27(0.12)  & \phn\textminus 5.2(\phn5.2)                  & I 7591            \\
      2002ew  & \nodata    & \nodata    & \textminus0.53(0.10) & \nodata                     &       \phs0.54(0.20)  &     \textminus10.2(10.2)                     & I 7964            \\
      2002gd  & \nodata    & 2.28(0.02) & \textminus0.49(0.05) & \nodata                     &       \phs0.46(0.13)  & \phn\textminus 0.6(\phn0.0)                  & I 7986, 7990      \\
      2002gw  & \nodata    & 2.45(0.03) & \textminus0.57(0.05) & 0.41(0.19)                  &       \phs0.61(0.14)  & \phn\textminus 2.3(\phn5.0)\tablenotemark{e} & I 7995, A14 \\
      2002hj  & \nodata    & 2.31(0.04) & \textminus0.34(0.07) & 0.21(0.31)                  &       \phs0.21(0.13)  & \phn\textminus 5.7(\phn5.7)                  & I 8006            \\
      2002hx  & \nodata    & 1.45(0.02) & \textminus0.48(0.09) & 0.18(0.25)                  &       \phs0.46(0.18)  & \phn\textminus 8.0(\phn8.0)                  & I 8015            \\
      2003B   & 0.56(0.01) & 2.66(0.03) & \textminus0.29(0.05) & 0.01(0.25)                  &       \phs0.12(0.10)  &     \textminus29.5(11.0)\tablenotemark{e}    & I 8042, A14 \\
      2003E   & \nodata    & 2.31(0.02) & \textminus0.74(0.08) & 1.11(0.31)                  &       \phs0.91(0.22)  &     \textminus11.3(\phn7.0)\tablenotemark{e} & I 8044, A14       \\
      2003T   & \nodata    & 2.21(0.03) & \textminus0.44(0.06) & 0.54(0.31)                  &       \phs0.38(0.14)  &     \textminus10.0(10.0)                     & I 8058            \\
      2003Z   & 0.62(0.02) & 2.01(0.04) & \textminus0.41(0.07) & \nodata                     &       \phs0.32(0.15)  & \phn\textminus 4.5(\phn4.5)                  & I 8062            \\
      2003bl  & \nodata    & 1.99(0.04) & \textminus0.36(0.09) & 0.01(0.16)                  &       \phs0.24(0.17)  & \phn\textminus 2.5(\phn3.0)\tablenotemark{e} & I 8086, A14       \\
      2003bn  & \nodata    & 2.37(0.03) & \textminus0.22(0.04) & 0.12(0.16)                  & \textminus0.01(0.08)  & \phn\textminus 6.1(\phn0.7)                  & I 8088            \\
      2003ci  & \nodata    & 1.98(0.04) & \textminus0.31(0.09) & 0.46(0.31)                  &       \phs0.14(0.16)  & \phn\textminus 8.0(\phn8.0)                  & I 8097            \\
      2003cn  & \nodata    & 1.85(0.03) & \textminus0.34(0.05) & 0.01(0.25)                  &       \phs0.19(0.11)  &     \textminus11.0(11.0)                     & I 8101            \\
      2003ef  & \nodata    & 1.79(0.02) & \textminus0.68(0.10) & 1.26(0.25)                  &       \phs0.79(0.23)  &     \textminus21.0(\phn0.0)                  & I 8131, 8132      \\
      2003ej  & \nodata    & 2.03(0.04) & \textminus0.53(0.08) & \nodata                     &       \phs0.54(0.17)  & \phn\textminus 4.5(\phn4.5)                  & I 8134            \\
      2003fb  & 0.77(0.01) & 1.95(0.03) & \textminus1.04(0.14) & 0.45(0.31)                  &       \phs1.44(0.36)  &     \textminus20.5(\phn6.0)\tablenotemark{e} & I 8143, A14       \\
      2003gd  & 0.99(0.02) & 1.11(0.02) & \textminus0.48(0.09) & 0.43(0.31)                  &       \phs0.45(0.19)  &     \textminus50.0(30.2)\tablenotemark{e}    & I 8150, H08       \\
      2003hg  & 0.63(0.06) & 2.47(0.01) & \textminus1.07(0.09) & \nodata                     &       \phs1.49(0.33)  & \phn\textminus 4.5(\phn4.5)                  & I 8184            \\
      2003hk  & 0.71(0.01) & 1.85(0.03) & \textminus0.42(0.09) & 0.67(0.31)                  &       \phs0.34(0.17)  & \phn\textminus 5.1(\phn4.0)\tablenotemark{e} & I 8184, A14       \\
      2003hl  & 0.66(0.01) & 1.92(0.01) & \textminus0.79(0.09) & 1.27(0.25)                  &       \phs1.00(0.24)  & \phn\textminus 4.5(\phn4.5)                  & I 8184            \\
      2003hn  & 0.66(0.01) & 2.19(0.02) & \textminus0.58(0.06) & 0.60(0.25)                  &       \phs0.62(0.16)  &     \textminus10.3(10.3)                     & I 8186            \\
      2003ho  & 0.86(0.01) & 1.02(0.02) & \textminus1.68(0.19) & 1.26(0.31)                  &       \phs2.57(0.59)  &     \textminus10.9(10.9)                     & I 8186            \\
      2003ib  & 0.60(0.02) & 1.70(0.03) & \textminus0.67(0.08) & \nodata                     &       \phs0.78(0.20)  &     \textminus10.0(\phn5.0)                  & I 8201            \\
      2003ip  & 0.79(0.01) & 1.82(0.02) & \textminus0.43(0.08) & 0.43(0.31)                  &       \phs0.37(0.17)  &     \textminus18.3(\phn4.0)\tablenotemark{e} & I 8214, A14       \\
      2003iq  & 0.57(0.01) & 2.21(0.03) & \textminus0.39(0.08) & 0.40(0.16)                  &       \phs0.29(0.16)  & \phn\textminus 1.5(\phn1.5)                  & I 8219            \\
      2004A   & 0.41(0.02) & 2.60(0.04) & \textminus0.48(0.07) & \nodata                     &       \phs0.45(0.16)  & \phn\textminus 6.9(\phn6.9)                  & I 8265            \\
      2004dj  & 0.51(0.01) & 2.33(0.03) & \textminus0.43(0.05) & 0.52(0.25)                  &       \phs0.36(0.12)  &     \textminus20.4(17.4)\tablenotemark{e}    & I 8377, H08       \\
      2004et  & 0.61(0.01) & 2.05(0.01) & \textminus0.37(0.03) & 0.15(0.25)                  &       \phs0.25(0.08)  & \phn\textminus 4.5(\phn0.5)                  & I 8413            \\
      2005ay  & 0.50(0.05) & 1.97(0.12) & \textminus0.28(0.06) & \nodata                     &       \phs0.10(0.11)  & \phn\textminus 3.7(\phn3.7)                  & I 8500, 8502      \\
      2005cs  & 0.67(0.01) & 1.99(0.01) & \textminus0.37(0.03) & 0.14(0.16)\tablenotemark{f} &       \phs0.25(0.07)  & \phn\textminus 1.5(\phn0.5)                  & I 8553, P09       \\
      2008in  & 0.49(0.01) & 2.22(0.02) & \textminus0.12(0.04) & \nodata                     & \textminus0.19(0.09)  & \phn\textminus 2.6(\phn0.2)                  & C 1636, R11       \\
      2009N   & 0.57(0.01) & 2.16(0.02) & \textminus0.67(0.05) & \nodata                     &       \phs0.79(0.17)  &     \textminus10.9(10.9)                     & C 1670            \\
      2009bw  & 0.61(0.01) & 2.50(0.02) & \textminus0.50(0.04) & \nodata                     &       \phs0.47(0.12)  & \phn\textminus 2.4(\phn2.4)                  & C 1743, I12       \\
      2009js  & 0.47(0.02) & 2.23(0.05) & \textminus0.38(0.05) & \nodata                     &       \phs0.27(0.11)  & \phn\textminus 5.5(\phn5.5)                  & C 1969            \\
      2009md  & 0.65(0.03) & 1.63(0.07) & \textminus0.54(0.06) & \nodata                     &       \phs0.55(0.15)  & \phn\textminus 7.9(\phn7.9)                  & C 2065, F11       \\
      2012A   & 0.64(0.01) & 1.93(0.02) & \textminus0.40(0.03) & \nodata                     &       \phs0.30(0.08)  & \phn\textminus 4.7(\phn4.7)                  & C 2974            \\
      2012aw  & 0.59(0.01) & 1.91(0.02) & \textminus0.37(0.04) & \nodata                     &       \phs0.26(0.09)  & \phn\textminus 0.8(\phn0.8)                  & C 3054, A 3996 \\
      2013ej  & 0.63(0.01) & 1.93(0.03) & \textminus0.23(0.03) & \nodata                     &       \phs0.01(0.06)  & \phn\textminus 1.6(\phn0.3)                  & C 3606, A 5237 \\\hline\\[-1.2cm]
        \tablecomments{\scriptsize Column 1: SN names. Columns 2 and 3: C3 slopes in the \{$\vri$\} and \{$\bvi$\} filter subset respectively. Column 4: C3 y-intercept for the \{$\bvi$\} filter subset. Columns 5 and 6: reddenings from spectroscopic analysis and from the C3($\bvi$) method respectively. Columns 7 and 8: midpoint epoch and references respectively.}
        \tablenotetext{a}{\scriptsize Using a fixed slope $m^{_{BV\!I}}_0=2.14\pm0.27$.}
        \tablenotetext{b}{\scriptsize Reddenings from \citet{Olivares_etal2010}, unless otherwise noted, adapted to the calibration of \citet{Schlafly_Finkbeiner2011} and assuming $R_V=3.1$.}
        \tablenotetext{c}{\scriptsize Epochs are expressed with respect to the discovery time. The shock breakout epoch lies in the range $t_m-\Delta t<t_0<t_m+\Delta t$.}
        \tablenotetext{d}{\scriptsize I: IAU Circular; C: IAU's Central Bureau for Astronomical Telegram; A: Astronomer's Telegram; H08: \citealt{Harutyunyan_etal2008}; P09: \citealt{Pastorello_etal2009}; F11: \citealt{Fraser_etal2011}; R11: \citealt{Roy_etal2011}; I12: \citealt{Inserra_etal2012}; A14: \citealt{Anderson_etal2014}.}
        \tablenotetext{e}{\scriptsize Unhelpful $t_n$.}
        \tablenotetext{f}{\scriptsize \citet{Dessart_etal2008}.}
      \end{tabular}
    \end{minipage}
\end{table}

To obtain the values of $n^{_{BV\!I}}$ for each SN, we first correct photometry for reddening due to our Galaxy using a \citet{Fitzpatrick1999} reddening law and $R_V=3.1$ \citep{Schlafly_Finkbeiner2011}. Then we adjust a straight line keeping the slope $m_0^{_{BV\!I}}$ in the linear fitting. In order to include the uncertainty of $m_0^{_{BV\!I}}$ in the $n^{_{BV\!I}}$ uncertainty, we perform Monte Carlo simulations varying ramdomly the slope according to its uncertainty. Table \ref{ladj} lists the C3 y-intercept values for the SNe in the \{$\bvi$\} filter subset (Column 4). In order to set the zero point for the color excesses, we need to choose among our SNe those with smallest reddening according to the criterion given in Section \ref{host_galaxy_extinction}, i.e., those with the lowest value of $n$ if the slope of the reddening vector is larger than the C3 slope, or those with the highest value of $n$ if the slope of the reddening vector is smaller than the C3 slope. Using the top panel of Figure \ref{REDD_LAW_vs_Rv} we can confirm that $m^{_{BV\!I}}_{\textbf{\textit{E}}}$ is smaller than $m^{_{BV\!I}}_0$ for $R_V\geq1$ with a high confidence level ($m^{_{BV\!I}}_{\textbf{\textit{E}}}<m^{_{BV\!I}}_0-3\sigma_{m^{_{BV\!I}}_0}$). In other words, for all the reasonable values of $R_V$, the reddening vector moves the C3 to the lower-right corner of Figure \ref{3g}, and therefore the SN with the highest value of $n^{_{BV\!I}}$ corresponds to the SN with the smallest reddening. Among our SNe, SN 1992af and SN 2008in have the highest values of $n^{_{BV\!I}}$. 
SN 2008in spectra show \ion{Na}{1} D absorption features at the redshift of its host galaxy, indicating a small reddening of $A_V=0.08$ mag \citep{Anderson_etal2014}, while for SN 1992af, although which we did not detect \ion{Na}{1} D lines, \citet{Olivares_etal2010} found a high amount of reddening through spectroscopic analysis (see Column 5 of Table \ref{ladj}). To determine if SN 1992af and SN 2008in can be use as a reddening zero point, we will analyze a group of four SNe with high and nearly equal values of $n^{_{BV\!I}}$, viz, SN 1993S, SN 1999eg, SN 2003bn, and SN 2013ej.

Among the four SNe found with possible small reddening, for SN 1993S and 2003bn we found no significant \ion{Na}{1} D interstellar lines in their spectra at the redshifts of their host galaxies. Similar result was found by \citet{Valenti_etal2014} for SN 2013ej. Thus these three SNe could be useful to define the zero point for reddening. In the case of SN 1999eg, the available spectra are fairly noisy, so we cannot identify or discard the presence of \ion{Na}{1} D lines. Therefore we do not use this SN to define the zero point for reddening. For the following analysis we adopt $n^{_{BV\!I}}_0=-0.23\pm0.02$ mag, i.e., the weighted mean of the $n^{_{BV\!I}}$ values for SN 1993S, SN 2003bn, and SN 2013ej.

Table \ref{ladj} summarizes previous reddening results obtained by \citet{Olivares_etal2010} from spectroscopic analysis $A_V$(spec) (Column 5, adapted to the calibration of \citealt{Schlafly_Finkbeiner2011}) and our C3($\bvi$) reddenings results $A_V(\bvi)$ (Column 6) for the SN set used in this work, using a CCM reddening law with $R_V=3.1$.

\begin{figure}[p]
  \centering
  \includegraphics[width=\columnwidth]{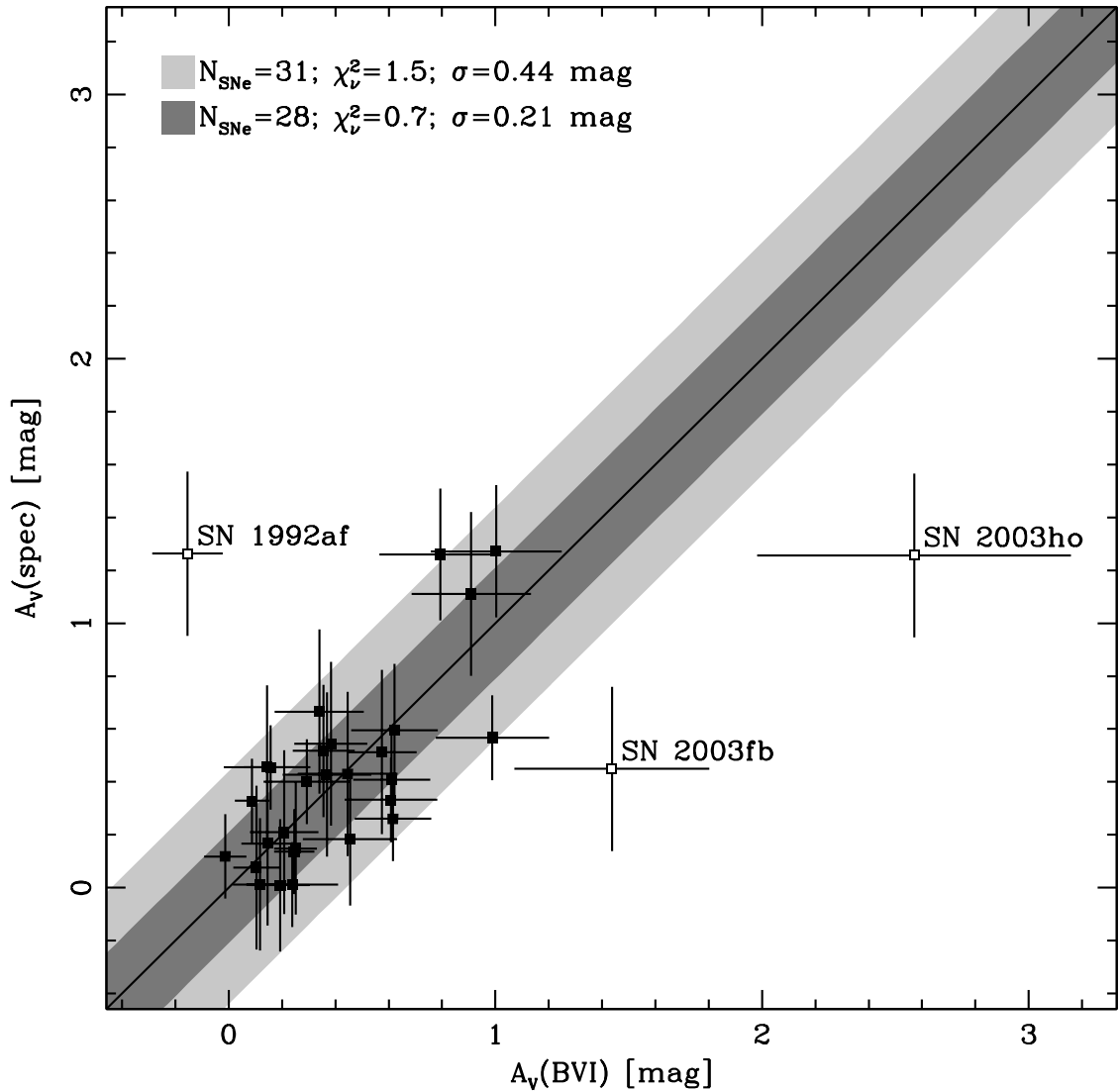}
  \caption{Comparison of $A_V$ obtained with the spectrum-fitting method and those obtained with the C3($\bvi$) method. Solid line represents the one-to-one relation. Open squares correspond to the outliers of the correlation. The shaded regions contain data within $1\sigma$ of the perfect correlation: using all SNe in our SN set (light grey region) and discarding the outliers (dark grey region). }
  \label{Av(SPEC)_vs_Av(BVI)}
\end{figure}

Figure \ref{Av(SPEC)_vs_Av(BVI)} shows the comparison between reddenings from spectroscopic analysis and from the C3($\bvi$) method. We see that both methods are well-correlated with the exception of SN 1992af, SN 2003fb, and SN 2003ho (empty squares), which increase the dispersion up to 0.44 mag. Our reddening measurement for SN 1992af, although negative, is consistent with zero reddening within $1.2\sigma$. Considering the non-detection of \ion{Na}{1} D lines, SN 1992af is more like a low-reddened than a high-reddened SN, as is suggested by spectroscopic analysis. When we include this SN to the group to define the reddening zero point, the value of $n^{_{BV\!I}}_0$ virtually does not change. For SN 2003ho we measure a \ion{Na}{1} D equivalent width (EW) of 1.5 \AA. Using the \citet{Poznanski_etal2012} relation and assuming $R_V=3.1$, the EW indicates $A_V\approx2.5$ mag, consistent with our reddening estimation. In general, differences between photometric and spectroscopic results for these three SNe can also be due to the use of late-time spectra (with flux correction in the case of SN 1992af and SN 2003fb), which are not the best for reddening determination with the spectroscopic analysis \citep{Olivares_etal2010}. Discarding the outliers, the dispersion is reduced to 0.21 mag, where the value of $\chi^2_\nu=0.7$ indicates that the dispersion is slightly lower than the combined uncertainties of both techniques. 

These results indicate that the C3($\bvi$) method can be used as a good reddening estimator, with SN 1992af, SN 1993S, SN 2003bn, and SN 2013ej as reddening zero point, so we will use this method throughout this work. Although negative reddening values have not physically meaningful, negative values in our SN set are statistically consistent with zero reddening, with the exception of SN 2008in which differs by $2.1\sigma$ from zero reddening. A possible explanation for the negative value is the effect of a high metallicity, which reduces the flux in the $B$-band (via line blanketing) producing higher values of the $\bv$ color index. This effect moves the C3 upward, acting contrary to the reddening effect. Based on the correlation between metallicity and the strength of the \ion{Fe}{2} 5018 \AA\ absorption line displayed by theoretical models, \citet{Dessart_etal2014} derive a metallicity for SN 2008in of $\approx2Z_\sun$, which is the highest value in their SN set\footnote{\citet{Roy_etal2011}, however, infer a sub-solar metallicity for the region of the SN occurrence.}. So it is possible that SN 2008in be a low or zero reddened SN, with $n^{_{BV\!I}}>n^{_{BV\!I}}_0$ due to a high metallicity. Moreover, \citet{Dessart_etal2014} find SN 2003bn, one of our SNe used to set the reddening zero point, among the SNe with the lowest metallicities in their SN set. So SN 2003bn, which is consistent with zero reddening, also works as a zero point for the metallicity effect. As the $n^{_{BV\!I}}$ value of SN 2003bn is virtually equal to $n^{_{BV\!I}}_0$, we will keep this value as the reddening zero point, but we will set $A_V=0$ mag for SN 2008in.

\subsection{Photospheric magnitude for SNe II}
Since we are trying to understand the intrinsic dispersion of the color-based standardization, we do not want to introduce large dispersion from the distances in our trial SN sample. So, in this work, we calibrate the photospheric magnitude using SNe with host galaxy distances measured with Cepheids. Among SNe with shock breakout epoch uncertainty of a few days, we found four SNe, viz., SN 1999em, SN 1999gi, SN 2005ay, and SN 2012aw with Cepheid distances. For SN 2005ay host galaxy (NGC 3938) we adopt the Cepheid distance for NGC 3982 which, like NGC 3938, is a member of the Ursa Mayor Group. In the case of SN 1999gi, the distance of its host galaxy (NGC 3184) is estimated through Cepheid distances for NGC 3319 and NGC 3198, which \citet{Tully1988} catalogued in a small group of four galaxies, although NGC 3184 is a galaxy relatively isolated \citep{Leonard_etal2002b}. Table \ref{SN_mu} shows the name and distance modulus of the host galaxy of the four aforementioned SNe II. We call this set ``nearby SNe''.
\begin{table}[ht]\footnotesize
    \caption{Host Galaxy Distance Moduli of the Nearby SNe}
    \label{SN_mu}
    \begin{minipage}{\columnwidth}
      \centering
      \renewcommand{\arraystretch}{0.667}
      \begin{tabular}{l l c}\hline\hline
        SN Name & Host Galaxy & $\mu(\sigma)$                \\
                &             & (mag)                        \\\hline
         1999em & NGC 1637    & 30.40(0.07)                  \\
         1999gi & NGC 3184    & 30.74(0.08)\tablenotemark{a} \\
                &             & 30.80(0.08)\tablenotemark{b} \\
         2005ay & NGC 3938    & 31.87(0.15)\tablenotemark{c} \\
         2012aw & NGC 3351    & 30.10(0.07)                  \\\hline
     \tablecomments{Values from \citet{Saha_etal2006}.}
     \tablenotetext{a}{From Cepheid distance for NGC 3319, a possible member of a small group of four galaxies, that contains to NGC 3184 \citep{Tully1988}.}
     \tablenotetext{b}{From Cepheid distance for NGC 3198. Similar to the case of NGC 3319.}
     \tablenotetext{c}{From Cepheid distance for NGC 3982, member of the Ursa Mayor Group.}
      \end{tabular}
    \end{minipage}
\end{table} 

With the knowledge of the crucial parameters for our nearby SNe (Tables \ref{ladj} and \ref{SN_mu}), we construct the photospheric magnitude according to Equation \ref{Mph_M}. For this, we need photometric and spectroscopic data at the same epoch. We choose to interpolate photometric points rather than spectroscopic data because in our SN set the former is better sampled. In order to do the photometric interpolation, we first realize fits to the light curves using the \textit{Local Polynomial Regression Fitting} \citep{Cleveland_etal1992} which performs a polynomial regression over small local intervals along the domain using a routine called \texttt{loess}. With this we interpolate with cubic-spline, using the routine \texttt{splint} \citep{Press_etal1992}. 

For SNe II-P \citet{Poznanski_etal2009} and \citet{Olivares_etal2010} found values of $R_V=1.5\pm0.5$ and $R_V=1.4\pm0.1$ respectively, lower than the standard Galactic value of 3.1. Hereafter we will adopt $R_V=1.5$ as the representative value for SNe II.

\subsubsection{Color-magnitude diagram}
\begin{figure}[p]
  \centering
  \includegraphics[width=\columnwidth]{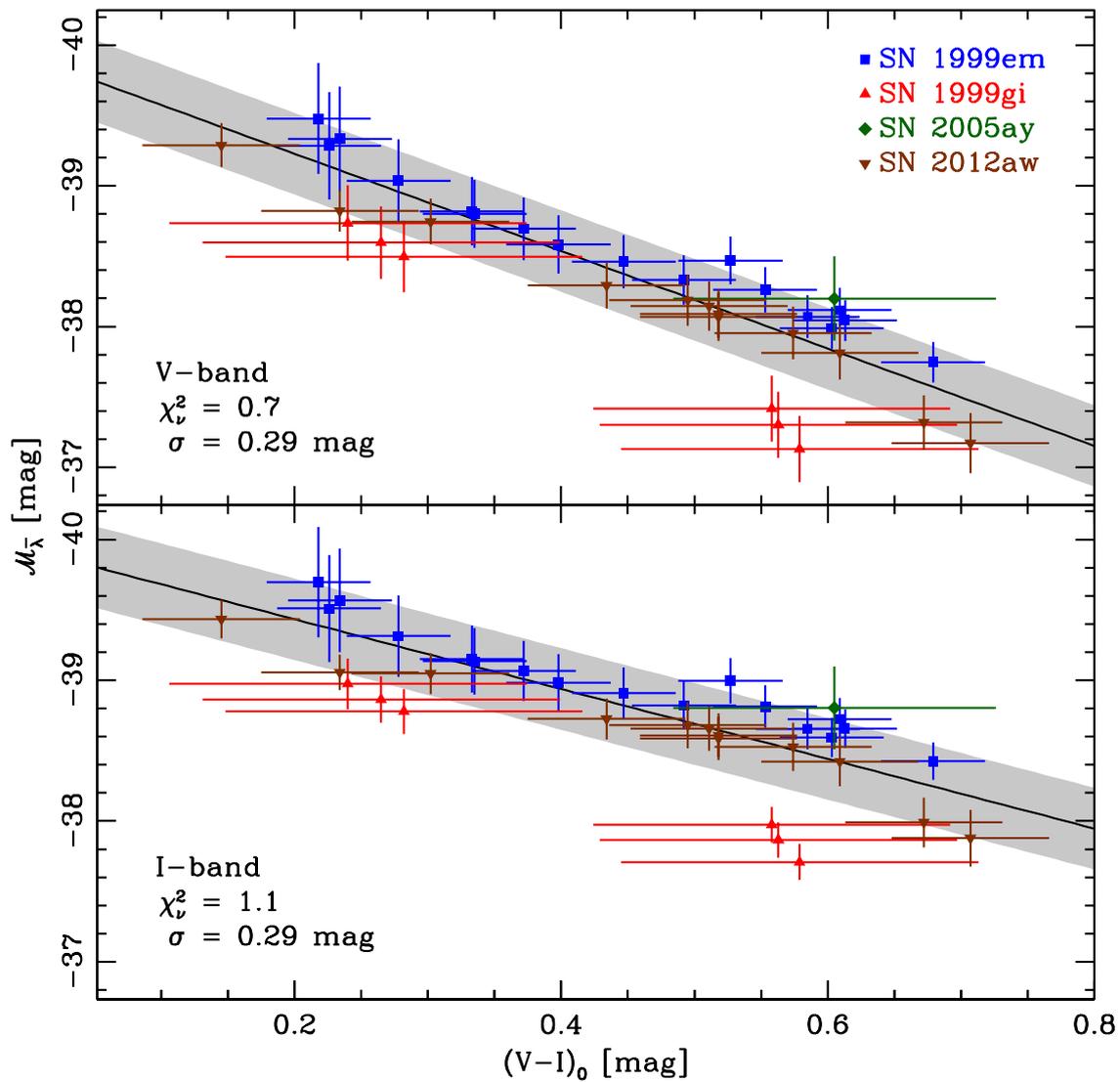}
  \caption{Photospheric magnitude versus corrected $\vi$ color for our nearby SNe, using $V$ (top) and $I$ (bottom) bands. The gray shading corresponds to the 1$\sigma$ dispersion.}
  \label{VIvsVI_mu}
\end{figure}

Figure \ref{VIvsVI_mu} shows the evolution of $\mathcal{M}_V$ (top) and $\mathcal{M}_I$ (bottom) with $\vi$ color for our nearby SNe throughout their photospheric phase.
The sequences are best represented by straight lines, i.e.,
\begin{equation}
 \mathcal{M}_{\overline{\lambda}}=a_1+a_2(\vi)_0,
\end{equation}
The parameters of the linear fit are listed in Table \ref{PARAMETERS}. The values of $\chi^2_\nu\lesssim1$ indicate that the intrinsic dispersion is undetectable at the current level of precision. Ignoring the observational uncertainties, the dispersion of 0.29 mag represents a relative distance scatter of 13\%.

\subsubsection{Photospheric light curve}
For each SN II the photospheric magnitude is a time-dependent quantity, which is given implicitly in Equation \ref{Mph_M} by the $i$ subindex. In that case points describe some sort of light curve where, in addition to being corrected by distance, as in the case of an absolute light curve, they are also corrected by the $\mathcal{R}$-term. We can see in Figure \ref{VIvsVI_mu} that time enters as an independent variable in both axes. In the case of $\mathcal{M}_{\overline\lambda,i}$ through the $\mathcal{R}$-term (this is an inference from theory). In the case of the $(\vi)_0$ color, from its evolution with time (this is an empirical inference). We then remove the factor $t-t_0$ included in the $\mathcal{R}$-term, i.e.,
\begin{equation}\label{ISCM}
 \mathcal{M}_{\overline\lambda,i} - 5\log{\left(\frac{t-t_0}{100\text{ d}}\right)}= m^\text{corr}_{\overline\lambda,i}-A_h(\overline\lambda)-\mu-\mathcal{R}^*_i\equiv\mathcal{M}^*_{\overline\lambda,i},
\end{equation}
where now
\begin{equation}\label{R*}
 \mathcal{R}^*_i\equiv5\log{\left(\frac{10\text{ pc}/100\text{ d}}{v_{\text{ph},i}}\right)}
\end{equation}
gives the correction by the size of the source, where time is given implicitly in the $i$ subindex. As we remove time from the y-axis, the uncertainty in this axis is reduced and becomes uncorrelated with the uncertainty of the new x-axis.

\begin{figure}[p]
  \centering
  \includegraphics[width=\columnwidth]{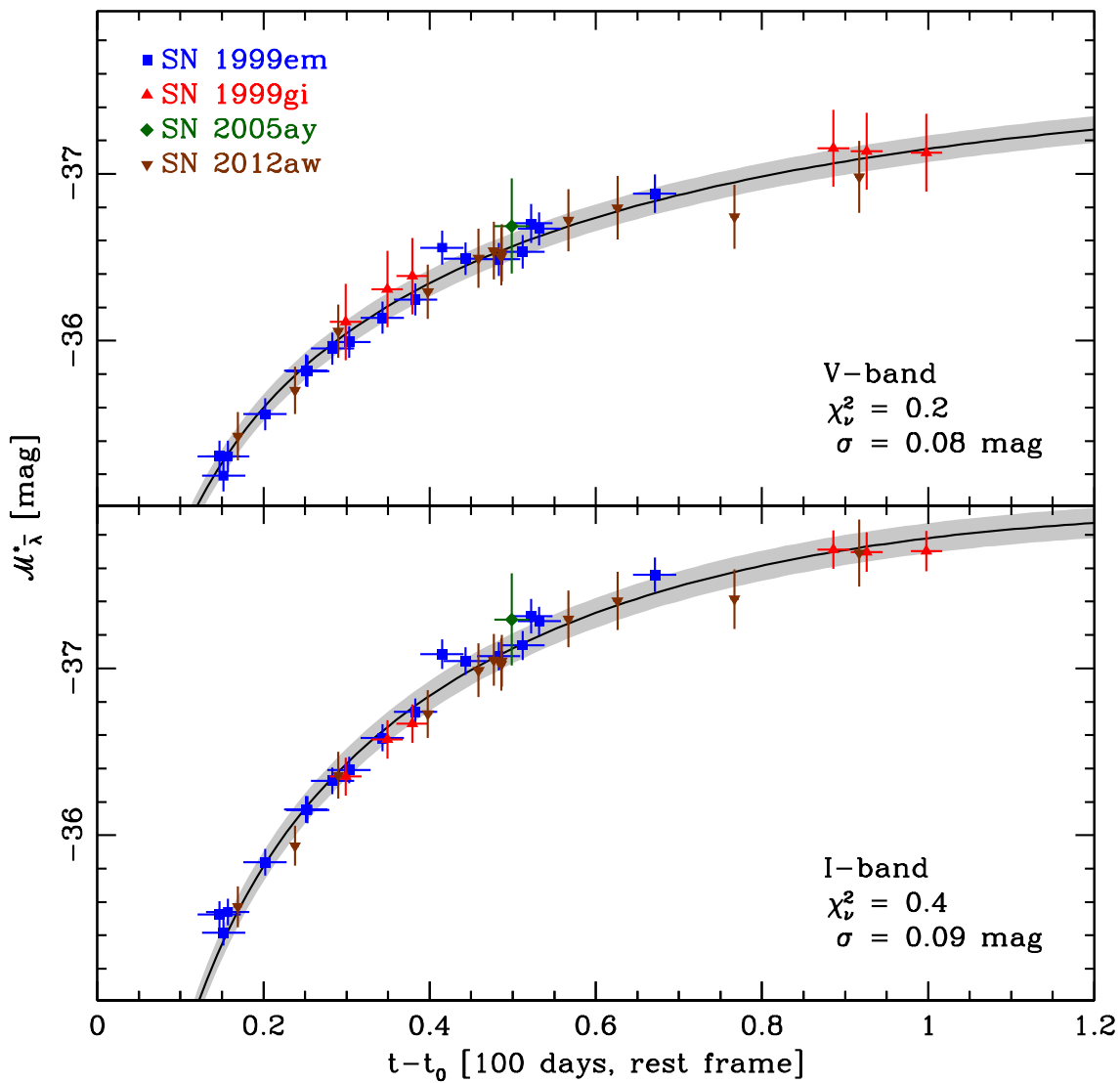}
  \caption{$\mathcal{M}^*_V$ and $\mathcal{M}^*_I$ evolution with time since shock breakout epoch for our nearby SNe, using $V$ (top) and $I$ (bottom) bands. The gray shading corresponds to the 1$\sigma$ dispersion.}
  \label{VIvsISCM_mu}
\end{figure}

Figure \ref{VIvsISCM_mu} shows the time evolution of $\mathcal{M}^*_V$ (top) and $\mathcal{M}^*_I$ (bottom) for our nearby SNe throughout their photospheric phase, with the shock breakout epoch as time reference.
Modeling $m^\text{corr}_{\overline\lambda}$ during the photospheric phase as a straight line and the photospheric velocity as a power law, i.e.,
\begin{equation}\label{power_law}
 v_{\text{ph}}=\alpha\left(\frac{t-t_0}{100\text{ d}}\right)^\beta
\end{equation}
\citep[e.g.,][]{Olivares_etal2010}, the function to fit will be
\begin{equation}\label{ISCM_POL}
 \mathcal{M}^*_{\overline\lambda}=a_1+a_2\left(\frac{t-t_0}{100\text{ d}}\right)+a_3\log{\left(\frac{t-t_0}{100\text{ d}}\right)}.
\end{equation}
The parameters of the fit are listed in Table \ref{PARAMETERS}. The mean dispersion of 0.09 mag implies relative distances with a precision of 4\%. 

\begin{table}[ht]\footnotesize
    \caption{Parameters for Photospheric Magnitude Fit}
    \label{PARAMETERS}
    \begin{minipage}{\columnwidth}
      \centering
      \renewcommand{\arraystretch}{0.667}
      \begin{tabular}{l c c c c}\hline\hline
         Adjust            &    $a_1(\sigma)$      &$a_2(\sigma)$\tablenotemark{a}&$a_3(\sigma)$         &$\sigma$\\
                           &         (mag)         &                              &     (mag)            & (mag)  \\\hline
$\mathcal{M}_{V}(\vi)$     & \textminus39.92(0.15) & 3.46(0.30)                   &      \nodata         & 0.29 \\
$\mathcal{M}_{I}(\vi)$     & \textminus39.92(0.15) & 2.46(0.32)                   &      \nodata         & 0.29 \\
$\mathcal{M}^*_{V}(t-t_0)$ & \textminus37.70(0.21) & 0.55(0.26)                   & \textminus2.85(0.23) & 0.08 \\
$\mathcal{M}^*_{I}(t-t_0)$ & \textminus38.99(0.27) & 1.21(0.31)                   & \textminus4.20(0.34) & 0.09 \\\hline
        \tablenotetext{a}{Without units for color evolution, and with units of mag for time evolution.}
      \end{tabular}
    \end{minipage}
\end{table} 

\begin{figure}[p]
  \centering
  \includegraphics[width=\columnwidth]{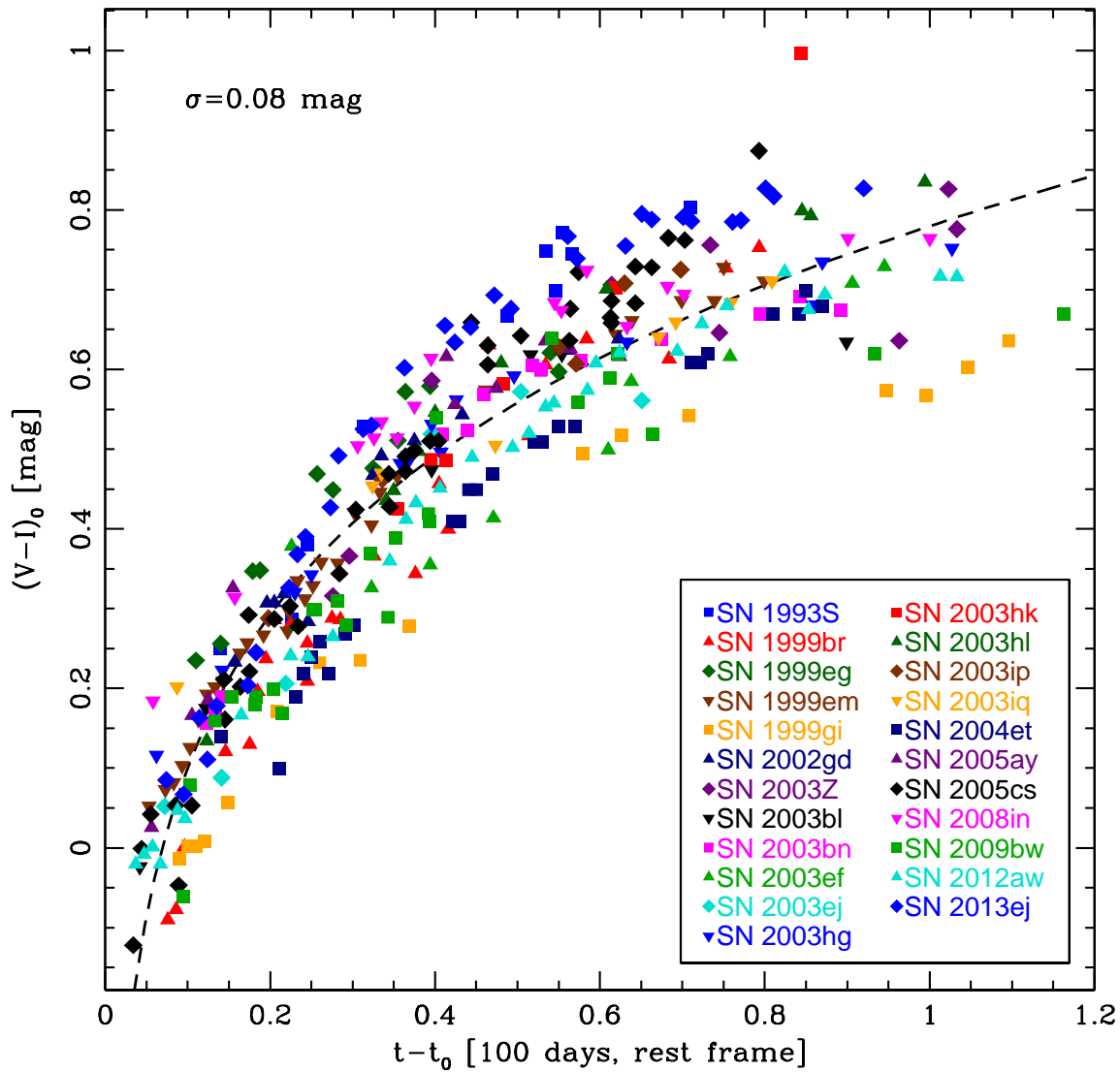}
  \caption{$\vi$ color curves in the rest frame of 23 SNe II during the photospheric phase, with shock breakout epoch uncertainty smaller than 4.5 days (the uncertainty for SN 1999em), corrected for Galactic and host galaxy extinction. The scatter of 0.08 mag with respect to the average color curve (dashed line) indicates that the assumption of a same color curve for SNe II would be inaccurate.}
  \label{COLOR_EVO}
\end{figure}

We see that, changing color by time, the dispersion is reduced. It is not an expected result because temperature (estimated through a color index) should be a variable more correlated with the SN flux than time. Switching the independent axis from color to time involves bringing in the relation between color and time, i.e., the color curve. However, even if our nearby SNe have the same color curve, points from Figure \ref{VIvsVI_mu} would be moved in the x-axis in the same way, and therefore it would show the same scatter than Figure \ref{VIvsISCM_mu}. Suppose that the good behavior of photospheric magnitude with time is a general result, with a characteristic dispersion of 0.09 mag. Figure \ref{COLOR_EVO} shows the $\vi$ color curves of 23 SNe in our set with shock breakout epoch uncertainty lower than 4.5 days (the uncertainty for SN 1999em), corrected by Galactic and host galaxy extinction. The scatter in the branch is principally due to differences between color curves of each SNe. So we can estimate an average color curve with a scatter of 0.08 mag. Replacing time by color in Figure \ref{VIvsISCM_mu}, the scatter in color produces an indirect contribution of $\sim0.24$ mag to the y-axis, producing a total dispersion of $\sim0.26$ mag or $12\%$ in relative distances. This result is consistent with the dispersion obtained with our nearby SNe (Figure \ref{VIvsVI_mu}) and with the typical scatter of $15\%$ in relative distances obtained with SN atmosphere models through the EPM \citep{Jones_etal2009}.

The small scatter found for the photospheric light curve, although it was obtained with four SNe only, gives us a hint about a possible homogeneity degree in the radiative process of SNe II during the photospheric phase. This conclusion could be debated considering the diversity of slopes and magnitudes of SN II light curves \citep[e.g., Figure 2 in][]{Anderson_etal2014}. Here the correction by the $\mathcal{R}$-term becomes crucial.

Similar to Equation \ref{Mph_M}, 
\begin{equation}\label{GOLD_EQ}
 M_{\overline\lambda,i}=\mathcal{M}^*_{\overline\lambda,i}+\mathcal{R}^*(v_{\text{ph},i}), 
\end{equation}
where the photospheric magnitude and the $\mathcal{R}$-term are given by Equations \ref{ISCM} and \ref{R*} respectively. Equation \ref{GOLD_EQ} reads: if the photospheric magnitude evolution is the same for all SNe II, then the shape of the light curve is defined mostly by the evolution of the $\mathcal{R}$-term, which in turn depends on the velocity evolution. With Equation \ref{GOLD_EQ} we can understand the diversity of the observed SN II light curves during the photospheric phase in terms of the diversity of velocity evolution:

1. The slope of the light curve is given by
\begin{equation}\label{GOLD_EQ2}
  \dot{M}_{\overline\lambda,i}=\dot{\mathcal{M}}^*_{\overline\lambda,i}+\frac{5}{\ln(10)}\frac{|\dot{v}_{\text{ph},i}|}{v_{\text{ph},i}},
\end{equation}
where dots denote temporal derivative. For example, SNe with less massive H envelopes (i.e., photospheres being not well-supported by the H recombination) have photospheres receding rapidly in mass coordinate, therefore from Equation \ref{GOLD_EQ2} they have steeper light curves than H rich SNe. The previous description is the usual one to distinguish between Plateau and Linear SNe II.

2. The difference between absolute magnitudes of two SNe is given by
\begin{equation}\label{GOLD_EQ3}
  M_{1,i}-M_{2,i}=5\log{\left(\frac{v_{\text{ph},2,i}}{v_{\text{ph},1,i}}\right)}.
\end{equation}
In the case of a typical SN II-P (as SN 2004et), $v_{ph}\sim4000$ km s$^{-1}$ at the middle of the plateau. There are, however, II-P events characterized by low expansion velocities and low luminosities, they are known as subluminous SNe II-P \citep{Pastorello_etal2004,Spiro_etal2014}. These events are characterized by expansion velocities of the order of $v_{ph}\sim1500$ km s$^{-1}$ at the middle of the plateau. From Equation \ref{GOLD_EQ3}, subluminous events should be $\sim2$ mag fainter than typical SNe II-P.

To test the above points, we need more than four well-observed SNe: our nearby SNe are neither Linear nor subluminous events. So we select from our set 14 SNe, of which four are Linear (SN 1991al, SN 2003ci, SN 2003hk, and SN 2003ip) and one is subluminous with low expansion velocities (SN 2003bl). All of them are within the Hubble flow, i.e., $cz_{\text{\tiny CMB}}>3000$ km s$^{-1}$, where $cz_{\text{\tiny CMB}}$ is the host galaxy redshift in the CMB frame, in order to compute absolute magnitudes using redshift-based distances. They also have at least three spectroscopic observations in order to perform a power law fit (Equation \ref{power_law}) to the expansion velocities, and estimate the $\mathcal{R}$-term for each photometric point.

\begin{figure}[p]
  \centering
  \includegraphics[width=\columnwidth]{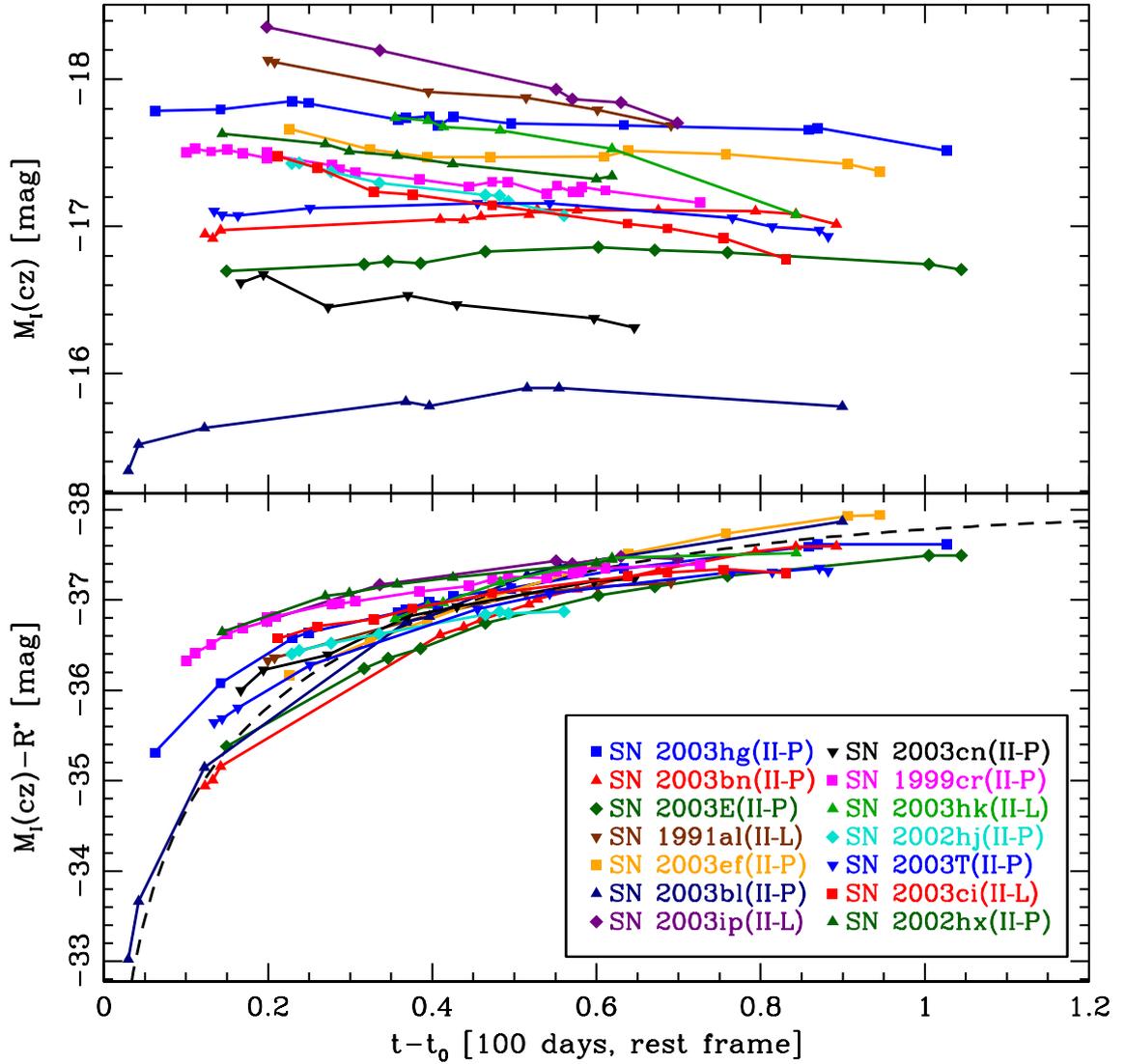}
  \caption{Top: $I$-band absolute light curves in the rest frame of 14 SNe II within the Hubble flow and with at least three spectroscopic observations. Bottom: Photospheric light curves for the same SNe, and the relation obtained from our nearby SNe (dashed line). SN names are sorted by increasing redshift.}
  \label{Iabs_Iph_z}
\end{figure}

The top panel of Figure \ref{Iabs_Iph_z} shows the $I$-band absolute light curves of 14 SNe II within the Hubble flow, using $H_0=72$ km s$^{-1}$ Mpc$^{-1}$. The spread between fainter and brighter events is of $\sim2.5$ mag. Bottom of Figure \ref{Iabs_Iph_z} shows the $I$-band photospheric light curves. We see that SNe are located approximately on the same locus than our nearby SNe (dashed line), independent if they are Plateau, Linear, or subluminous events. The largest dispersion occurs before $\sim40$ days, and is probably due to the uncertainty in the shock breakout epoch and the scarcity of early spectroscopic observations. In the range of 50--90 days the spread between fainter and brighter events is $\leq0.6$ mag. Figure \ref{Iabs_Iph_z} illustrates that the apparent diversity of slopes and magnitudes of SN II light curves during the photospheric phase is linked with the velocity profile of SNe.

\subsubsection{Photospheric magnitude method}
Working in Equation \ref{ISCM} with redshift instead of distance modulus, we obtain
\begin{equation}\label{SCM}
 m^\text{corr}_{\overline\lambda,i}-A_h(\overline\lambda)+5\log{(v_{\text{ph},i})}=5\log(cz)+\mathcal{M}^*_{\overline\lambda,i}-5\log{\left(H_0 100\text{ d}\right)},
\end{equation}
where $H_0$ is the Hubble constant. For an epoch $t_p$ at the middle of the plateau (e.g., 50 days since shock breakout), Equation \ref{SCM} reduces to the original version of the standardized candle method (SCM) proposed by \citet{Hamuy_Pinto2002}, i.e.,
\begin{equation}\label{SCM_HAMUY}
 m^\text{corr}_{\overline\lambda,p}-A_h(\overline\lambda)+\alpha\log{(v_{\text{ph},p})}=5\log(cz)+zp,
\end{equation}
where $\alpha$ and $zp$ are free parameters to fit from the observations, by minimizing the dispersion in the Hubble diagram. In Equation \ref{SCM} $\alpha=5$ as a consequence of the spherical symmetry assumption, and $zp=\mathcal{M}^*_{\overline\lambda,i}-5\log(H_0100\text{ d})$, which is constant for a fixed time. So, Equation \ref{ISCM} corresponds to the generalization of the SCM for various epochs throughout the photospheric phase. Hereafter we will refer to it as the photospheric magnitude method (PMM).

Most recent versions of the SCM \citep[e.g.,][]{Poznanski_etal2009,DAndrea_etal2010} replace in Equation \ref{SCM_HAMUY}
\begin{equation}\label{trampa}
 A_h(\overline\lambda)=R_{\overline\lambda,V\!-I}(R_V)\left[(\vi)_p-(\vi)_{0,p}\right],
\end{equation}
where $(\vi)_p$ is the color measured at $t_p$, and $(\vi)_{0,p}$ is the (supposed) intrinsic color at the same epoch. The reasons to use Equation \ref{trampa} in \ref{SCM_HAMUY} are: avoiding calculate the value of $(\vi)_{0,p}$ (essential for the reddening determination), which is absorbed by $zp$, and using the SCM to determine $R_V$ (i.e., a third parameter to adjust to the observations). In the case of the PMM, the incorporation of Equation \ref{trampa} in \ref{SCM} is inaccurate because it is equivalent to suppose that all SNe II have the same intrinsic color curve $(\vi)_{0,i}$. So, in order to use the PMM, we necessarily need the knowledge of the host galaxy extinction with the most appropriated value of $R_V$ for SNe II.

\subsubsection{Application: The Hubble constant}
For a SN with measured observables \{$m^{\text{corr}}_{\overline\lambda}$, $v_{\text{ph}}$\}, $A_h(\overline\lambda)$ and $t_0$, with photometric and spectroscopic data at the same epoch, we can calculate the distance modulus using the PMM via
\begin{equation}
  \mu_{\overline\lambda} = \langle m^{\text{corr}}_{\overline\lambda,i}-\mathcal{R}^*(v_{\text{ph},i})-\mathcal{M}^*_{\overline\lambda,i}(t_0)\rangle -A_h(\overline\lambda),
\end{equation}
\begin{equation}\label{s_mu}
  \sigma_{\mu_{\overline\lambda}}^2 = \sigma_{\langle m^{\text{corr}}_{\overline\lambda,i}-\mathcal{R}^*(v_{\text{ph},i})\rangle}^2 +\sigma_{A_h(\overline\lambda)}^2.
\end{equation}

\begin{table}[p]\scriptsize
    \caption{Hubble Diagram SNe Sample}
    \label{SN_muV_muI}
    \begin{minipage}{\columnwidth}
      \centering
      \renewcommand{\arraystretch}{0.667}
      \begin{tabular}{l c l c c c}\hline\hline
Host Galaxy\tablenotemark{a} &$cz_{\text{\tiny CMB}}$\tablenotemark{b} & SN Name  & SN Type\tablenotemark{c} & $\mu_V(\sigma)$&$\mu_I(\sigma)$\\
                          &    (km s$^{-1}$)                   &           &      &  (mag)      &   (mag)     \\\hline
NGC 6946                  & \phn\textminus141                  & 2004et & II-P & 28.65(0.08) & 28.73(0.05) \\
NGC 2403                  &       \phn\phs182                  & 2004dj & II-P & 27.68(0.11) & 27.79(0.05) \\
NGC 628                   &       \phn\phs359                  & 2003gd & II-P & 29.76(0.24) & 29.74(0.18) \\
                          &                                    & 2013ej & II-L & 30.01(0.07) & 29.88(0.05) \\
NGC 5194                  &       \phn\phs636                  & 2005cs & II-P & 29.77(0.08) & 29.63(0.05) \\
NGC 1637                  &       \phn\phs670                  & 1999em & II-P & 30.42(0.06) & 30.36(0.03) \\
NGC 3184                  &       \phn\phs831                  & 1999gi & II-P & 30.70(0.22) & 30.83(0.09) \\
NGC 6207                  &       \phn\phs862                  & 2004A  & II-P & 31.54(0.22) & 31.66(0.17) \\
NGC 3938                  &          \phs1038                  & 2005ay & II-P & 31.75(0.24) & 31.70(0.23) \\
UGC 2890                  &          \phs1077                  & 2009bw & II-P & 31.14(0.13) & 31.17(0.08) \\
NGC 3239                  &          \phs1084                  & 2012A  & II-P & 30.14(0.09) & 30.19(0.05) \\
NGC 1448                  &          \phs1102                  & 2003hn & II-P & 31.04(0.17) & 31.12(0.09) \\
NGC 1097                  &          \phs1105                  & 2003B  & II-P & 31.98(0.09) & 31.97(0.05) \\
NGC 3351                  &          \phs1127                  & 2012aw & II-P & 30.12(0.09) & 30.15(0.05) \\
NGC 2082                  &          \phs1246                  & 1992ba & II-P & 31.49(0.11) & 31.47(0.07) \\
NGC 918                   &          \phs1261                  & 2009js & II-P & 31.74(0.21) & 31.62(0.18) \\
NGC 4900                  &          \phs1285                  & 1999br & II-P & 32.11(0.15) & 32.11(0.07) \\
NGC 4487                  &          \phs1386                  & 2009N  & II-P & 31.77(0.19) & 31.83(0.10) \\
NGC 2742                  &          \phs1405                  & 2003Z  & II-P & 32.25(0.18) & 32.22(0.12) \\
NGC 5921                  &          \phs1646                  & 2001X  & II-P & 31.85(0.22) & 31.96(0.20) \\
NGC 3389                  &          \phs1656                  & 2009md & II-P & 32.15(0.24) & 32.25(0.19) \\
NGC 4303                  &          \phs1913                  & 2008in & II-P & 31.58(0.10) & 31.43(0.07) \\
NGC 772                   &          \phs2198                  & 2003hl & II-P & 32.46(0.25) & 32.42(0.11) \\
                          &                                    & 2003iq & II-P & 32.74(0.15) & 32.72(0.07) \\
NGC 7537                  &          \phs2304                  & 2002gd & II-P & 32.67(0.13) & 32.63(0.06) \\
NGC 922                   &          \phs2879                  & 2002gw & II-P & 33.27(0.15) & 33.34(0.06) \\
NGC 3120                  &          \phs3108                  & 1999ca & II-L & 32.97(0.10) & 32.88(0.07) \\
NGC 7771                  &          \phs3921                  & 2003hg & II-P & 33.52(0.34) & 33.51(0.14) \\
ESO 235--G58              &          \phs4134                  & 2003ho & II-P & 33.71(0.60) & 33.99(0.26) \\
2MASX J10023529--2110531  &          \phs4172                  & 2003bn & II-P & 34.07(0.07) & 34.03(0.04) \\
MCG --04--12--004         &          \phs4380                  & 2003E  & II-P & 34.05(0.23) & 34.05(0.10) \\
anon                      &          \phs4484\tablenotemark{d} & 1991al & II-L & 33.70(0.17) & 33.96(0.07) \\
NGC 4708                  &          \phs4504                  & 2003ef & II-P & 33.87(0.22) & 33.96(0.10) \\
NGC 5374                  &          \phs4652                  & 2003bl & II-P & 34.00(0.16) & 33.99(0.09) \\
UGC 11522                 &          \phs4996                  & 2003fb & II-P & 34.47(0.37) & 34.52(0.16) \\
UGC 327                   &          \phs5050                  & 2003ip & II-L & 33.79(0.17) & 33.74(0.10) \\
ESO 340--G38              &          \phs5359                  & 1992af & II-P & 34.08(0.13) & 33.97(0.07) \\
UGC 7820                  &          \phs5431                  & 2003ej & II-L & 34.45(0.19) & 34.42(0.12) \\
IC 849                    &          \phs5753                  & 2003cn & II-P & 34.34(0.16) & 34.31(0.13) \\
ESO 576--G34              &          \phs6363                  & 1999cr & II-P & 34.07(0.14) & 34.12(0.06) \\
IC 1861                   &          \phs6494                  & 1999eg & II-L & 34.97(0.14) & 34.77(0.13) \\
NGC 1085                  &          \phs6568                  & 2003hk & II-L & 34.80(0.16) & 34.79(0.08) \\
NPM1G +04.0097            &          \phs6869                  & 2002hj & II-P & 34.85(0.12) & 34.82(0.05) \\
MCG --04--48--015         &          \phs7203                  & 2003ib & II-P & 34.78(0.23) & 35.03(0.14) \\
UGC 4864                  &          \phs8662                  & 2003T  & II-P & 35.65(0.15) & 35.58(0.10) \\
NEAT J205430.50--000822.0 &          \phs8671                  & 2002ew & II-L & 35.32(0.20) & 35.33(0.11) \\
2MASX J07391822--6203095  &          \phs8908                  & 1993A  & II-P & 35.69(0.11) & 35.62(0.08) \\
UGC 6212                  &          \phs9468                  & 2003ci & II-L & 35.81(0.16) & 35.51(0.09) \\
2MASX J08273975--1446551  &          \phs9573                  & 2002hx & II-P & 35.50(0.17) & 35.38(0.08) \\
2MASX J22522390--4018432  &          \phs9645                  & 1993S  & II-L & 35.62(0.10) & 35.48(0.06) \\\hline\\[-1.2cm]
      \tablecomments{Distance moduli are computed using the midpoint time as the shock breakout epoch. Uncertainties do not include the intrinsic scatter of the PMM.}
      \tablenotetext{a}{Sorted by increasing redshift.}
      \tablenotetext{b}{Values from the NASA/IPAC Extragalactic Database, unless otherwise noted, with an error of 187 km s$^{-1}$ due to the uncertainty in the determination of the Local Group velocity.}
      \tablenotetext{c}{Old SN subclassification based on the $B$-band decline rate ($\beta_{100}^B$) criterion given in \citet{Patat_etal1994}.}
      \tablenotetext{d}{\citet{Hamuy2001}.}
      \end{tabular}
    \end{minipage}
\end{table}
\begin{figure}[p]
  \centering
  \includegraphics[width=\columnwidth]{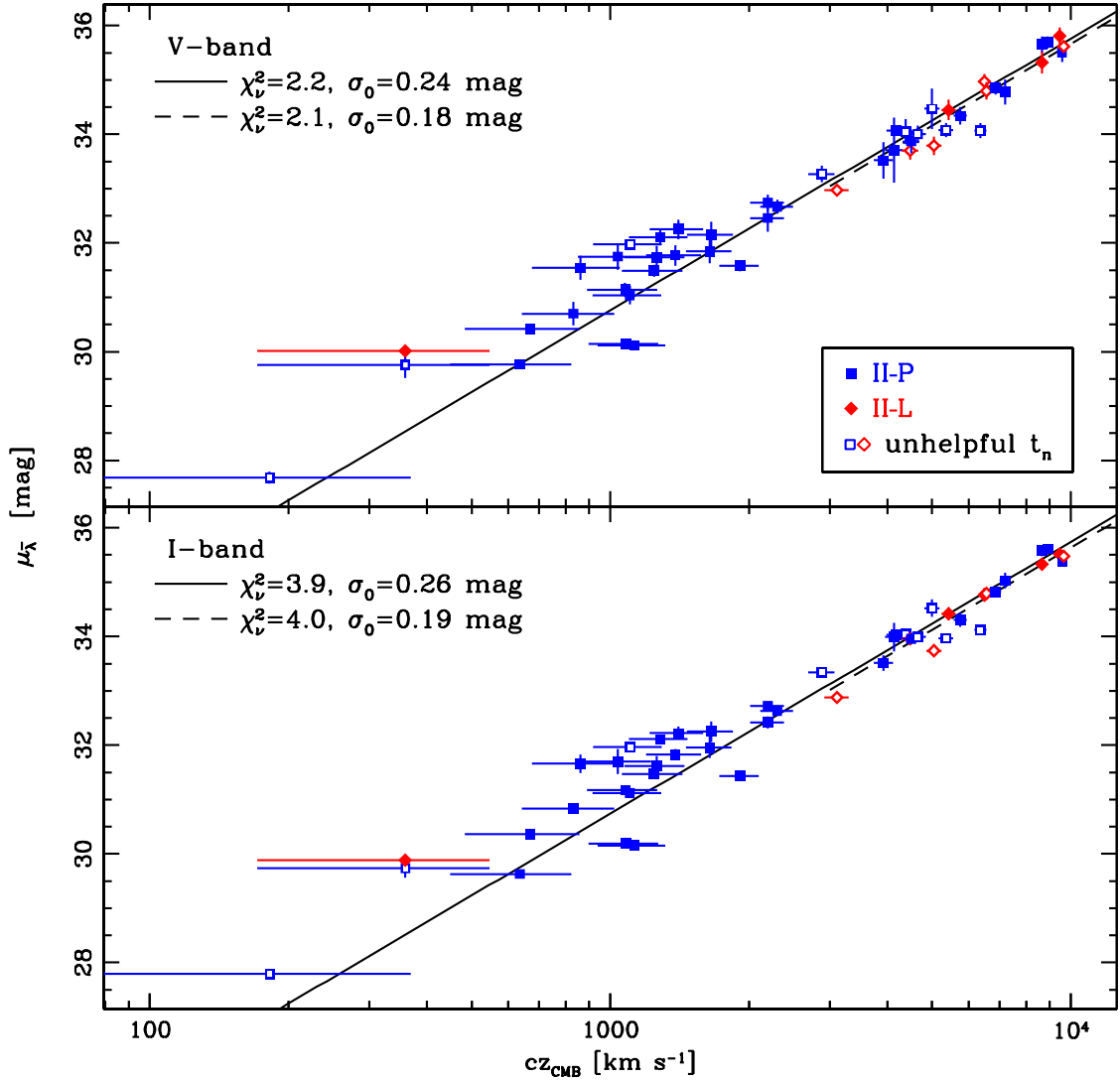}
  \caption{Hubble diagram for our SN set, using parameters derived in this work. Solid lines represent linear fits using all SN in our set, and dashed lines represent linear fits using only SNe with $cz_{\text{\tiny CMB}}>3000$ km s$^{-1}$.}
  \label{HUBBLE_DIA}
\end{figure}

Table \ref{SN_muV_muI} shows our results for the $V$ and $I$ bands, where we use the midpoint time as the shock breakout epoch. A direct way to test the previous results is through the construction of a Hubble diagram. Figure \ref{HUBBLE_DIA} shows the Hubble diagram using $V$ (top) and $I$ (bottom) bands. To obtain the Hubble constant and the intrinsic dispersion $\sigma_0$, we maximize the marginal likelihood function $\mathcal{L}$ minimizing the quantity
\begin{equation}
 -2\ln\mathcal{L}=\sum_{\text{SN}}\left\{\ln V+\frac{\left(\mu-\left(25-5\log H_0\right)-5\log cz_{_\text{\tiny CMB}}\right)^2}{V}\right\},
\end{equation}
\begin{equation}
 V = \sigma_{\mu}^2+\left(\frac{5}{\ln 10}\frac{\sigma_{cz_{_\text{\tiny CMB}}}}{cz_{_\text{\tiny CMB}}}\right)^2+\sigma_0^2
\end{equation}
\citep[e.g.,][]{Kelly2007}, where the summation is over all the SNe in our sample. For the $V$ and $I$ bands we obtain an intrinsic dispersion of 0.24--0.26 mag respectively, indicating a relative distance scatter of 12\%. Using only SNe within the Hubble flow, the intrinsic scatter is reduced to
0.16--0.19 mag, indicating a relative distance scatter of 7--9\%.

\begin{figure}[p]
  \centering
  \includegraphics[width=\columnwidth]{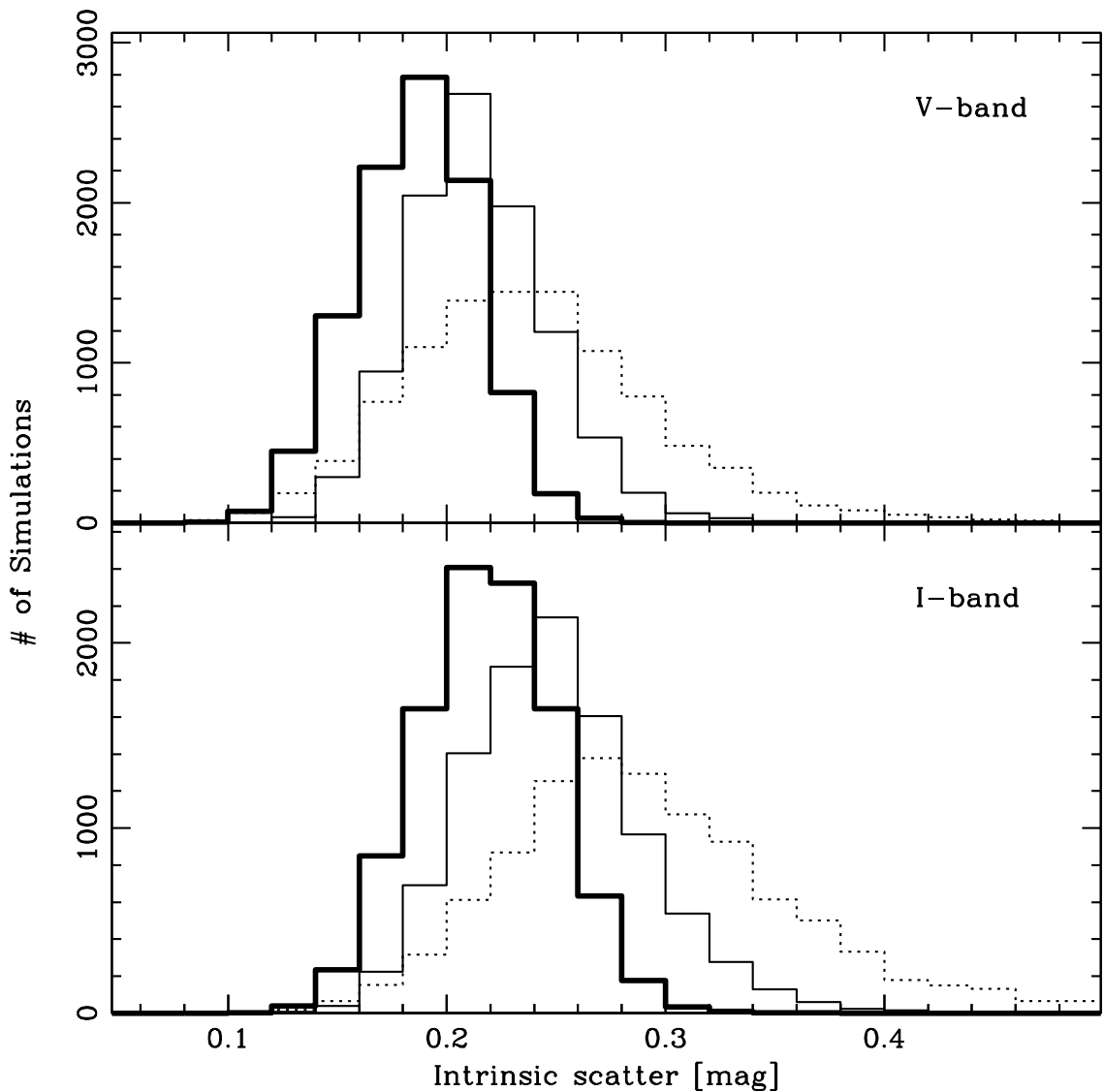}
  \caption{Histrograms showing the intrinsic scatter in the $V$ (top) and $I$ (bottom) bands, performing Monte Carlo simulations for the shock breakout epoch in the ranges $t_m-\Delta t<t_0<t_m+\Delta t $ (dotted line), $t_m<t_0<t_m+\Delta t$ (thin line), and $t_m-\Delta t<t_0<t_m$ (thick line).}
  \label{RMS_hist}
\end{figure}

The mean uncertainty in the shock breakout epoch for our 24 SNe in the Hubble flow is of 6 days, so the choice of $t_0$ may have an important effect on the intrinsic scatter. Figure \ref{RMS_hist} shows the results of Monte Carlo simulations for $t_0$ in the range $t_m-\Delta t<t_0<t_m+\Delta t$ (dotted line). The intrinsic scatter is around 0.24 and 0.28 mag for $V$ (top) and $I$ (bottom) bands respectively, values greater than our previous results. To study the contribution of the shock breakout epoch uncertainty to the intrinsic scatter, we divide the range where $t_0$ lies in two: $t_m-\Delta t<t_0<t_m$ (lower range) and $t_m<t_0<t_m+\Delta t$ (upper range). Figure \ref{RMS_hist} shows the results of Monte Carlo simulations for $t_0$ in the lower range (thick line) and in the upper range (thin line). Both ranges yield lower values for the intrinsic scatter than the values obtained with the entire range. However, the best results are obtained with the lower limit, with intrinsic scatter of 0.19--0.22 mag. This may be because, in general, our 24 SNe in the Hubble flow were first detected after maximum light, so the shock breakout occurred, in fact, several days before the first detection. So choosing the lower limit can correct for this systematic error.

The similarity between the results obtained with $t_m$ as $t_0$, and those obtained with $t_0$ in the range $t_m-\Delta t<t_0<t_m$ means than an uncertainty of 3 days in the shock breakout epoch does not produce an important contribution to the intrinsic scatter. 

We repeat the previous analysis using the 13 SNe in the Hubble flow with shock breakout epoch estimated with nondetection. With the aforementioned correction of the shock breakout epoch, we obtain a mean intrinsic scatter of 0.12 mag, or 6\% in relative distances. Also we obtain $H_0=68$--69 km s$^{-1}$ Mpc$^{-1}$. This value is based on the calibration obtained with our nearby SNe (Figure \ref{VIvsISCM_mu}), which in turn is based on the \citet{Saha_etal2006} calibration ($\mu_{_\text{LMC}}=18.54$ mag).

\section{Discussion}\label{discussion}

\subsection{The C3 method}
Reddening measurements through C3($\bvi$) method have a systematic uncertainty, mostly due to the line blanketing, that affects SNe II $B$-band after maximum light. So, despite the good agreement between reddening from the C3($\bvi$) method and from the spectroscopic analysis, it will be necessary to take into account the effect that metallicity has over reddening measurements in order not to propagate this effect to the $V$ and $I$ bands. Other source of uncertainty is the difference between photometric systems, so future works will need to consider the application of the $S$-correction \citep{Stritzinger_etal2002}.

\subsection{SBM for SNe II}\label{SBM_PMM}
The SBM has been used successfully in the Cepheid distance measurements with the subsequent calibration of the Cepheid period-luminosity relation \citep[e.g.,][]{Storm_etal2011}. The calibration of the SBM is made possible by the current ability to measure angular diameters of Galactic Cepheids through interferometric observations \citep{Welch_1994,Kervella_etal2004a}. In the case of SNe II, interferometric observations are possible only for the nearest events through VLBI, while indirect measurements of the angular sizes require the knowledge of expansion velocities, the shock breakout epoch, and the distance to the SN. The complex scenario for SNe II was probably the reason why astronomers chose to use methods based on models, as the EPM, instead of calibrating the SBM for SNe II.

\citet{Kasen_Woosley2009} suggest that the SBM for SNe II is used through the SCM, arguing that the temperature in the middle of the plateau is nearly constant. In addition, using SN II models, they find a relation between absolute magnitude and photospheric velocity during the plateau phase similar to our Equation \ref{ISCM_POL}, in which we include a linear term to model Linear events. We remark that the SBM relates the surface brightness with a color index and not with time. So the result obtained by \citet{Kasen_Woosley2009} is the first approach at a time-based standardization for SNe II, being our work an empirical proof.

\subsection{The future of the PMM}\label{PMM_next}
Based on our results, the PMM can replace the SCM, the latter being only a special case. This is an important improvement because, unlike the SCM, we will have not one but several points for each SN, depending on the number of observations we have and the interpolation we use. So we can obtain more accurate results as more photometric and spectroscopic data are available. In our case, the observational uncertainty (first term of Equation \ref{s_mu}) is reduced with the number of spectroscopic observations. Also with more photometric observation we can reduce the host galaxy extinction uncertainty (second term of Equation \ref{s_mu}). 

In order to reduce the uncertainty produced by the shock breakout epoch estimation, we need constrain it in a range of one week. This requirement was previously achieved in SN samples used by \citet{Poznanski_etal2009} and \citet{DAndrea_etal2010}. It will be important that future surveys for SNe II take into account this constrain.

Based on the resulting dispersion, our work shows that the most promising method to measure distances to SNe II is not the SBM (Figure \ref{VIvsVI_mu}) but the PMM (Figures \ref{VIvsISCM_mu} and \ref{Iabs_Iph_z}). In a future work we will investigate the possible application of the SBM and the PMM to other SN classes during their photospheric phase.

\section{Conclusions}\label{conclusions}
In this work we test improved ways of estimating distances using SNe II. We find a temperature-dependent quantity that we call ``photospheric magnitude'', which is equivalent to the surface brightness previously used for Cepheid distance measurements through its relation with a suitable color index. To study the applicability of the method, we use 50 SNe II with redshifts lower than $10^4$ km s$^{-1}$, and with photometric and spectroscopic data during the photospheric phase. We also develop a method to measure host galaxy extinctions, equivalent to that proposed by \citet{Natali_etal1994} for open clusters. Our main conclusion are the following:

\begin{list}{}{}
 \item[1.] $\bv$ versus $\vi$ color-color curves provide us with the possibility to measure reddening. The comparison between reddenings from spectroscopic analysis and from the C3($\bvi$) method is satisfactory within the errors of both techniques.
 
 \item[2.] We construct a CMD using our nearby SNe. All of them are in galaxies with distances measured with Cepheids, and with uncertainty in the shock breakout epoch smaller than a few days. The CMD shows a linear relation, with a dispersion of 0.29 mag, that represents a scatter in relative distances of 13\%.
 
 \item[3.] Using time since shock breakout instead of color as independent variable, the above relation turns into a photospheric light curve. We identify it as the generalization of the SCM for various epochs throughout the photospheric phase. The dispersion of 0.09 mag represents a scatter in relative distances of 4\%.

 \item[4.] The diversity of slopes and magnitudes of SN II light curves during the photospheric phase is mostly produced by the expansion velocity evolution of each SN.
  
 \item[5.] Finally, we applied the time-based standardization to our SN set in order to calculate their distances and to construct a Hubble diagram. Using only SNe within the Hubble flow with well-constrained shock breakout epoch, we obtain $H_0=68$--69 km s$^{-1}$ Mpc$^{-1}$. The mean intrinsic dispersion of 0.12 mag represents a scatter in relative distances of 6\%. It confirms the low intrinsic dispersion of the method.
 
\end{list}

We acknowledge support by projects IC120009 ``Millennium Institute of Astrophysics (MAS)'', P10-064-F ``Millennium Center for Supernova Science'' of the Iniciativa Cient\'ifica Milenio del Ministerio Econom\'ia, Fomento y Turismo de Chile, and by projects FONDAP 15010003, ALMA-CONICYT 31110008, and BASAL PFB-06 of CONICYT. This research has made use of the NASA/IPAC Extragalactic Database (NED) which is operated by the Jet Propulsion Laboratory, California Institute of Technology, under contract with the National Aeronautics and Space Administration. This work has made use of the Weizmann Interactive Supernova Data Repository (\url{http://www.weizmann.ac.il/astrophysics/wiserep}). A preliminary version of this work \citep{Rodriguez_2013} was developed as a Master Thesis.

\end{document}